\documentclass[twocolumn]{aastex7}

\usepackage{graphicx}
\usepackage{apjfonts}
\usepackage{hyperref}
\usepackage{subfigure}

\submitjournal{ApJ}

\shorttitle{Bifrost models of the Quiet Sun. I. Comparison with solar observations} 

\shortauthors{Go\v{s}i\'{c} et al.}

\begin{document}

\newcommand{\vhhtxt}[1]{\textcolor{cyan}{#1}}
\newcommand{\vhhcom}[1]{\footnote{\textcolor{blue}{#1}}}

\newcommand{\FeIion}{{\ion{Fe}{1}}}
\newcommand{\FeIHMIline}{{\FeIion~$6173$~\AA\/}}
\newcommand{\FeIsixthreeooneline}{{\FeIion~$6300.1$~\AA\/}}
\newcommand{\FeIsixthreeotwoline}{{\FeIion~$6300.2$~\AA\/}}
\newcommand{\CaIIK}{{\ion{Ca}{2} K}}
\newcommand{\CaIIIR}{{\ion{Ca}{2}~$8542$~\AA\/}}
\newcommand{\MgIIk}{{\ion{Mg}{2} k}}
\newcommand{\MgIItriplet}{{\ion{Mg}{2}~2799~\AA\/}}
\newcommand{\SiIV}{{\ion{Si}{4}~$1393$~\AA\/}}
\newcommand{\FeIX}{{\ion{Fe}{9}~$171$~\AA\/}}
\newcommand{\FeXV}{{\ion{Fe}{15}~$284$~\AA\/}}
	
\title{Bifrost models of the Quiet Sun. I. Comparison with solar observations}

\correspondingauthor{M.~Go\v{s}i\'{c}}
\author[0000-0002-5879-4371]{M.~Go\v{s}i\'{c}}
\affiliation{Lockheed Martin Solar \& Astrophysics Laboratory,
3251 Hanover St, Palo Alto, CA 94304, USA}
\affiliation{SETI Institute, 339 Bernardo Ave, Suite 200, Mountain View, CA 94043, USA}
\email[show]{mgosic.solar@gmail.com}

\author[0000-0003-0975-6659]{V. H. Hansteen}
\affiliation{Lockheed Martin Solar \& Astrophysics Laboratory,
3251 Hanover St, Palo Alto, CA 94304, USA}
\affiliation{SETI Institute, 339 Bernardo Ave, Suite 200, Mountain View, CA 94043, USA}
\affiliation{Rosseland Center for Solar Physics, University of Oslo, P.O. Box 1029 Blindern, N-0315 Oslo, Norway}
\affiliation{Institute of Theoretical Astrophysics, University of Oslo,
P.O. Box 1029 Blindern, N-0315 Oslo, Norway}
\email{v.h.hansteen@astro.uio.no}

\author[0000-0002-3234-3070]{A.~Sainz Dalda}
\affiliation{Lockheed Martin Solar \& Astrophysics Laboratory,
3251 Hanover St, Palo Alto, CA 94304, USA}
\affiliation{SETI Institute, 339 Bernardo Ave, Suite 200, Mountain View, CA 94043, USA}
\email{asainz.solarphysics@gmail.com}

\author[0000-0002-8370-952X]{B.~De Pontieu}
\affiliation{Lockheed Martin Solar \& Astrophysics Laboratory,
3251 Hanover St, Palo Alto, CA 94304, USA}
\affiliation{Rosseland Center for Solar Physics, University of Oslo, P.O. Box 1029 Blindern, N-0315 Oslo, Norway}
\affiliation{Institute of Theoretical Astrophysics, University of Oslo, P.O. Box 1029 Blindern, N-0315 Oslo, Norway}
\email{bdp@lmsal.com}

\author[0000-0003-2088-028X]{L. H. M. Rouppe van der Voort}
\affiliation{Rosseland Center for Solar Physics, University of Oslo, P.O. Box 1029 Blindern, N-0315 Oslo, Norway}
\affiliation{Institute of Theoretical Astrophysics, University of Oslo, P.O. Box 1029 Blindern, N-0315 Oslo, Norway}
\email{l.r.van.der.voort@astro.uio.no}

\begin{abstract}
Studying the emergence of magnetic fields is essential for understanding the physical mechanisms behind various phenomena in the solar atmosphere. Most importantly, the emerging fields offer valuable insights into how energy and mass are transferred to the upper solar atmosphere. As a result, they have garnered significant attention from both observational and theoretical perspectives. In this article, we present two models of quiet-Sun-like magnetic fields generated by the Bifrost code. We compare these models with observations from the Swedish 1-meter Solar Telescope (SST) and the Interface Region Imaging Spectrograph (IRIS). By tracking the magnetic features in both the SST and Bifrost data, we determine the similarities and differences between the fields identified in the models and those observed. We conduct a quantitative comparison of various properties, such as flux content, flux densities, horizontal and line-of-sight velocities, lifetimes, sizes, and surface interactions. Additionally, we identify and analyze the properties of the largest emerging bipoles in the SST and Bifrost data. Our findings indicate that the magnetic bipoles in the Bifrost simulations are generally stronger than those observed with the SST. However, a qualitative comparison of the chromospheric and transition region responses to the emerging fields in the Bifrost models, SST, and IRIS observations shows similar heating processes occurring above and around the emerging fields. Finally, we outline our plans for future work aimed at studying the emergence of magnetic fields in the quiet Sun, with a particular focus on the chromosphere and upper atmospheric layers.
\end{abstract}

\keywords{Sun: magnetic field -- Sun: photosphere -- Sun: chromosphere -- Sun: transition region}

\section{Introduction}
\label{introduction}

Solar magnetic fields have long been recognized as significant drivers of various solar phenomena that greatly impact our space environment. These fields are organized across a wide range of spatial and temporal scales, from large-scale active regions (ARs) to ephemeral and network (NW) structures, down to the smallest detectable internetwork (IN) magnetic elements. 

In ARs, solar magnetic fields can generate energetic events such as solar flares and coronal mass ejections, which are closely related to geomagnetic storms \citep{2021LRSP...18....4T}. In addition to those large-scale events, numerous studies have shown that magnetic fields outside ARs, referred to as the quiet Sun (QS), play an important role in maintaining solar magnetism \citep{2019LRSP...16....1B}, can produce small-scale eruptions and jets, and locally heat the solar atmosphere \citep{2020ApJ...894..104P, 2021ApJ...921L..20P, 2021A&A...656L...4B, 2023ApJ...943L..14M}. This often happens while the magnetic field ascends through the solar atmosphere and interacts with the ambient magnetic structures \citep{2009ApJ...700.1391M, 2020A&A...633A..67K, 2021ApJ...911...41G}. Thus, it is critically important to study how the magnetic fields of the QS are generated and transported through the solar interior and atmosphere and how they facilitate energy and mass transfer. Addressing these questions may allow us to decipher how the solar dynamo operates and on what scales. Additionally, doing so could shed light on how the upper solar atmosphere is heated to millions of degrees.

The QS features strong kilogauss (kG) magnetic fields located at the borders of supergranular cells, forming what is known as the NW \citep{1969SoPh....9..347S}. The NW fields are generated when magnetic flux is transported by horizontal flows from the inner parts of supergranular cells to their edges \citep{2012ApJ...758L..38O, 2013ApJ...770L..36G, 2014ApJ...797...49G, 2018ApJ...854..118A}. This process causes magnetic flux to accumulate at the borders, resulting in the formation of NW elements that appear as concentrated, discrete patches of magnetic flux, with field strengths in the kG range \citep[e.g.,][]{1987A&A...188..183S, 
1996A&A...315..610G, 2014ApJ...789....6R, 2015ApJ...810...79R}.

Between the NW fields, magnetic flux rises to the surface on supergranular scales, in the form of bipolar features known as ephemeral regions (ERs). They are termed ``ephemeral'' due to their short lifetimes, typically lasting around 3-4 hr \citep{1973SoPh...32..389H, 2000RSPTA.358..657T, 2001ApJ...555..448H}. ERs migrate toward the NW regions, where they interact with existing flux patches or directly create NW structures \citep{1997ApJ...487..424S}. The average unsigned flux of ERs follows a power-law distribution that gradually decreases toward the larger structures characteristic of ARs \citep{2009ApJ...698...75P, 2011SoPh..269...13T}. This property, combined with indications that the emergence rate of ERs does not fluctuate in sync with the solar cycle, led researchers to conclude that ERs may originate from both the dispersal of ARs and local mechanisms operating on smaller scales \citep{2003ApJ...584.1107H}.

At the smallest temporal and spatial scales are the IN fields, which are small, weak, and short-lived magnetic concentrations \citep[e.g.,][]{1985AuJPh..38..961Z, 1995SoPh..160..277W, 2010SoPh..267...63Z, 2012ApJ...751....2O, 2012ApJ...746..182O, 2022ApJ...925..188G}. These fields reside in the interiors of supergranular cells and appear in the photosphere in the form of magnetic bipoles \citep{2002ApJ...569..474D,2007ApJ...666L.137C, 2009ApJ...700.1391M, 2010A&A...511A..14G, 2012ApJ...745..160G, 2012A&A...537A..21P}, magnetic sheets \citep{2019A&A...622L..12F} or unipolar fields \citep{2008ApJ...674..520L, 2010ApJ...720.1405L, 2008A&A...481L..33O}. 

Different theoretical models have been proposed to explain the origin of IN fields: they could be generated by a local dynamo at or very close to the solar surface \citep{1999ApJ...515L..39C, 2001A&G....42c..18C, 2007A&A...465L..43V, 2010A&A...513A...1D, 2014ApJ...789..132R}, flux recycling from decaying ARs \citep{2001ASPC..236..363P, 2005A&A...441.1183D}, flux emergence from deeper layers, e.g., from a deep convection zone or an overshoot-layer dynamo as an extension of ARs and ERs \citep{2009SSRv..144..275D}, or by shallow recirculation in granular convection \citep{2018ApJ...859L..26M,2018ApJ...859..161R, 2020ApJ...903L..10F}.

Once IN fields appear in the photosphere, they can continue rising through the solar atmosphere. Similar to AR small-scale loops \citep{2014ApJ...781..126O, 2016ApJ...825...93O, 2014ApJ...794..140V, 2015ApJ...810..145D}, observations show that a fraction of IN fields may reach the chromosphere and even the transition region (TR). However, this seems insufficient to sustain the global heating of the chromospheric layers, let alone that of the corona \citep{2024ApJ...964..175G}. Regarding the magnetohydrodynamic (MHD) simulations, some suggest that the emergence of IN magnetic fields in the chromosphere is difficult, due to a lack of magnetic buoyancy \citep{2018ApJ...859L..26M, 2019ApJ...878...40M}, which could potentially be mitigated by ion-neutral coupling \citep[see, e.g.,][]{2017Sci...356.1269M, 2020ApJ...889...95M}. In contrast, other MHD models \citep{2008ApJ...679L..57I, 2015Natur.522..188A} indicate that IN magnetic loops are created and emerge as a result of a local dynamo and can reach the chromosphere and impact the higher layers of the solar atmosphere through heating and small-scale eruptions driven by magnetic reconnection. 

The origins of QS magnetic fields continue to be a topic of ongoing debate among researchers, with various hypotheses proposed to explain their formation. It is equally unclear which mechanisms are responsible for the upward propagation of these magnetic fields through the solar atmosphere. Understanding these dynamics is crucial for comprehending solar magnetism, including phenomena such as solar flares and coronal mass ejections, which can have significant effects on space weather and, consequently, on Earth. Further investigation into magnetic field generation and its ascent may shed light on the physical processes that govern solar activity.

With this work, we are initiating a project aimed at resolving the discrepancies between these models and determining the factors that drive the emergence of QS magnetic fields throughout the solar atmosphere, as well as exploring how these fields contribute to chromospheric heating.

In this first paper of the series, we will present Bifrost models of NW- and IN-like magnetic structures, which will be utilized in our future studies as part of this project. The models display multiple events of emerging flux, which we compare against high-resolution solar observations obtained with the Swedish 1 m Solar Telescope \citep[SST;][]{2003SPIE.4853..341S} and the Interface Region Imaging Spectrograph \citep[IRIS;][]{2014SoPh..289.2733D}. The structure of the paper is as follows. Section~\ref{data} introduces the observational data used. Section~\ref{tracking} describes how we identified and tracked magnetic elements over time. In Section~\ref{inversions_sst}, we explain how we obtained the magnetic and thermodynamic properties of the magnetic fields from both the Bifrost models and SST observations. Section~\ref{results} presents a comparison between the Bifrost models and SST observations, including examples of newly emerged magnetic fields. Finally, we conclude with a discussion and summary of our findings in Section~\ref{conclusions}.

\section{Observations and Bifrost models}
\label{data}

\begin{deluxetable*}{lccccc}
	\tablewidth{\textwidth}
	\tablecolumns{5} 
	\tablecaption{Description of the SST observations for the \ion{Fe}{1} 6173 \AA\/, \ion{Ca}{2} 8542 \AA\/ and H$\alpha$ 6563 \AA\/ spectral lines.\label{table1}} 
	\tablehead{ \colhead{Line (\AA\/)} & \colhead{$g_{\rm eff}$} & \colhead{$N_{\lambda}$} & \colhead{$\Delta\lambda$ (m\AA\/)} & \colhead{Wavelength Positions (m\AA\/)} & \colhead{Continuum Point (m\AA\/)}} \startdata
	\ion{Fe}{1} 6173 & 2.5 &  &  &  &  \\
        \hspace{1em} 2015 (full Stokes) &  & 12 & 35 & From $-240$ to $240$ & $315$\\
        \hspace{1em} 2016 (full Stokes) &  & 12 & 35 & From $-240$ to $240$ & $315$\\
    \ion{Ca}{2} 8542 & 1.1 &  &  &  \\
        \hspace{1em} 2015 (Stokes I) &  & 25 & 100 & From $-1200$ to $1200$ & $\pm$1200\\
        \hspace{1em} 2016 (Stokes I) &  & 11 & Varying & $-800$, $-400$, $-240$, $-110$, $0$, $110$, $240$, $400$, $800$ & $\pm$1750\\
        \hspace{1em} 2016 (full Stokes) &  & 2 &  & $-450$, $450$ &\\
    H$\alpha$ 6563 & 1.05 &  &  &  \\
    \hspace{1em} 2015 (Stokes I) &  & 15 & Varying & $-1500$, from $-1200$ to $1200$ at 200~m\AA\/ steps, 1500 & \\
	\enddata
    \tablecomments{For these spectral regions, we list their effective Land\'{e} $g$-factors ($g_{\rm eff}$), number of wavelengths ($N_{\lambda}$), wavelength step size ($\Delta\lambda$), wavelength positions concerning the corresponding resting line centers (m\AA\/), and continuum points. The continuum points listed for the \ion{Ca}{2} 8542 \AA\/ line are located in the far wings.}
\end{deluxetable*}

In this work, we employ QS observations obtained with SST and IRIS, which we compare with Bifrost simulations. These data sets show the spatiotemporal evolution of the QS magnetic fields in the photospheric layers and their propagation up to the upper chromosphere. Additionally, IRIS provides insights into the physical conditions of the plasma in the TR. Below, we present a detailed overview of these observations and simulations.

\subsection{SST observations}

We obtained two data sets from SST using the CRisp Imaging SpectroPolarimeter \citep[CRISP;][]{2008ApJ...689L..69S}. CRISP allows full Stokes measurements, which we carried out in the \ion{Fe}{1} 6173~\AA\/ (photosphere) and \ion{Ca}{2} 8542~\AA\/ (chromosphere) lines in QS regions around the disk center. In the H$\alpha$ 6563~\AA\/ line (upper chromosphere), CRISP operated in intensity-only mode. The CRISP instrument has a pixel size of $0\farcs057$, sufficient to critically sample the diffraction limit of $0\farcs16$ at 6300 \AA\/. The two recorded data sets have the following characteristics:

\begin{itemize}
    \item The data set taken on 2015 October 11 (SST15) starts at 08:44~UT and ends at 10:42~UT, monitoring QS at a viewing angle of $\mu=0.78$. The data include full Stokes measurements in the \ion{Fe}{1} 6173 \AA\/ and the \ion{Ca}{2} 8542 \AA\/ lines. Stokes-I-only measurements were taken in the \ion{Ca}{2} 8542 \AA\/ and H$\alpha$ 6563 \AA\/ lines. The cadence is about $\sim24.1$~s, with a total duration of 117 minutes for the time series.
    \item The measurements taken from 07:29~UT to 09:18~UT on 2016 August 3 (SST16) observed a QS area at the disk center ($\mu=0.99$) in the \ion{Fe}{1} 6173 \AA\/ (full Stokes vector) and the \ion{Ca}{2} 8542 \AA\/ (intensity-only mode) lines. Additionally, data were captured in the wings of the \ion{Ca}{2} 8542 line, recording all four Stokes parameters. The recordings have a cadence of 20~s and last for 109 minutes. No data were collected in the H$\alpha$ 6563 \AA\/ line.
\end{itemize}

Both data sets were reduced using the CRISPRED pipeline \citep{2015A&A...573A..40D}. To improve the image quality across the fields of view (FOVs), the images were restored by applying the multi-object, multi-frame blind-deconvolution technique \citep[MOMFBD;][]{2005SoPh..228..191V}. Residual seeing deformations were corrected as described by \cite{2012A&A...548A.114H}. The time sequences of restored spectral line scans were aligned and stretched, correcting for longer-timescale seeing variations \citep{1994ApJ...430..413S}. Finally, 5 minute and 3 minute oscillations from the photospheric and chromospheric images were removed using a subsonic filter \citep{1989ApJ...336..475T, 1992A&A...256..652S}.

\begin{figure*}[!t]
	\centering
	\resizebox{1\hsize}{!}{\includegraphics[width=1.0\textwidth, bb = -1016 -351 1660 1068, trim={2.5cm 2.5cm 2.5cm 2.5cm}, clip] {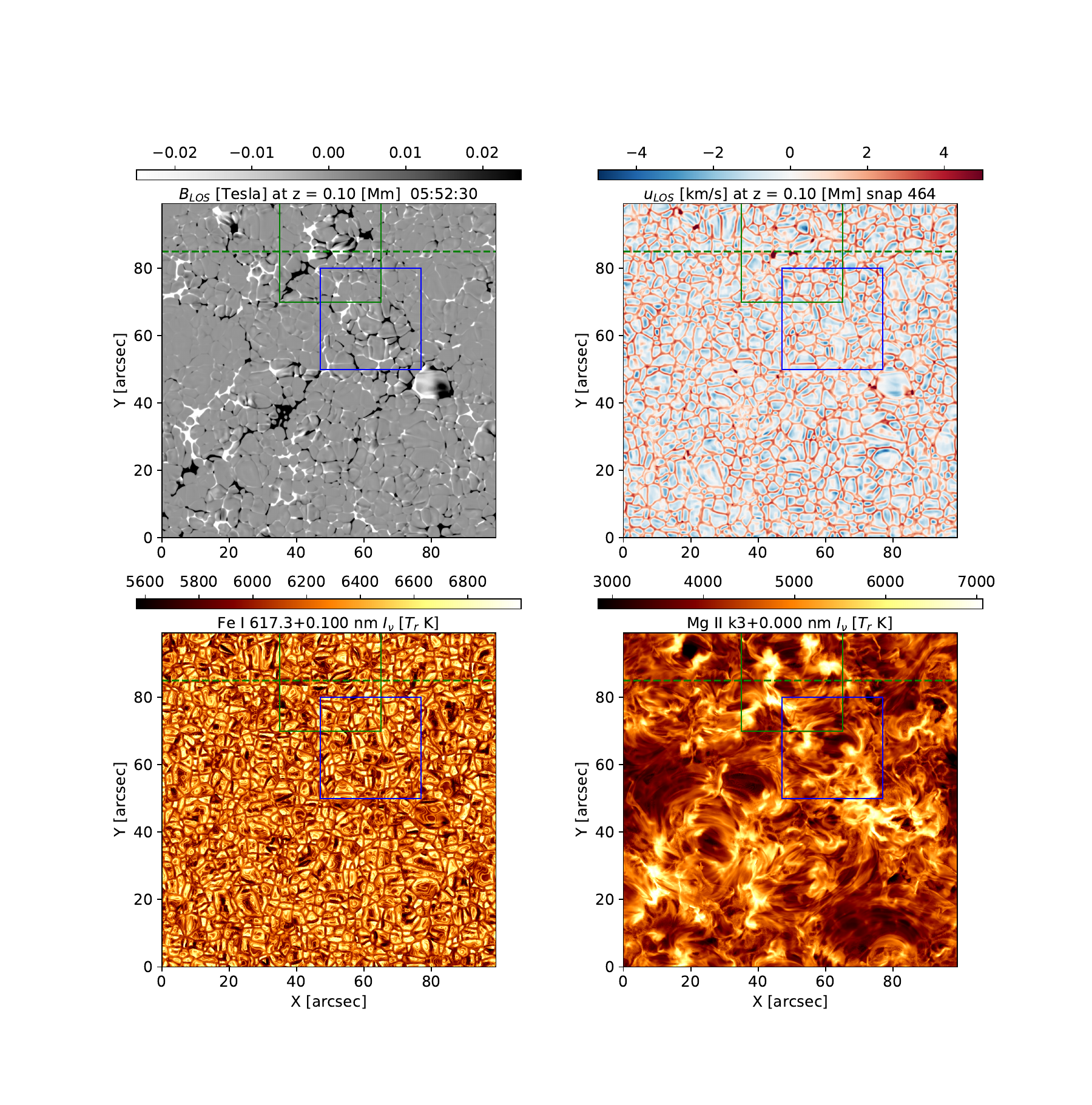}}
	\caption{Overview of the emerging flux model, showing the vertical component of the magnetic field and the vertical velocity 100~km above the photosphere (upper row) and the intensity in the red wing of the \FeIHMIline\ line and in the core of the \ion{Mg}{2}~k line (bottom row). The green rectangle shows the region ``Bifrost~1'' and the dashed green line is the location of the spectral line examples shown in later figures. The blue rectangle shows the location of the region ``Bifrost~2.'' The snapshots were generated in the {\tt nw072100} Bifrost simulation with horizontal spatial resolution of 100~km.}
	\label{fig:nw_overview_464}
\end{figure*}

The largest cluster of the newly emerging QS fields is detected in the SST15 data set, revealing the spatiotemporal evolution of an ER. The SST16 sequence, on the other hand, shows weaker fields and granular size loops. A summary of the observational sequences is provided in Table~\ref{table1}, organized by line. There, we include a list of all the wavelengths at which the measurements were carried out in each spectral line used. 

\subsection{IRIS observations} 

The SST measurements were coordinated with IRIS observations \citep{2020A&A...641A.146R}.
The SST15 data are complemented by IRIS observations recorded from 08:05:08~UT until 11:02:14~UT, resulting in an overlap of 2 hr between the two data sets. IRIS made medium-dense 32-step rasters, taking spectra in the near-ultraviolet (NUV) band from 2793 to 2806 \AA\/. Additionally, two spectral regions in the far-ultraviolet (FUV) domain were included, from 1332 to 1357 \AA\/ and from 1390 to 1406 \AA\/. This setup allows for sampling the solar atmosphere from the photosphere up to the TR. The cadence for the spectral observations was 3.1~s per raster step, leading to a total raster cadence of $\sim100$~s and covering a QS region of $11\times62$~arcsec$^{2}$, accounting for solar rotation. Slit-jaw images were taken using filters: \ion{C}{2}~1335~\AA\/ (SJI 1330), \ion{Si}{4}~1400~\AA\/ (SJI 1400), \ion{Mg}{2}~k~2796~\AA\/ (SJI 2796), and the \ion{Mg}{2}~h wing at 2832 \AA\/ (SJI 2832). The cadence is 12~s for all slit-jaw images, except for the SJI 2832 images, which were recorded at a cadence of 50~s. These IRIS measurements were spatially and spectrally binned on board by 2, covering an FOV of about $60\times68$~arcsec$^{2}$ (the pixel size is 0\farcs33).

\begin{figure*}[!t]
	\centering
	\resizebox{1\hsize}{!}{\includegraphics[width=1.0\textwidth]{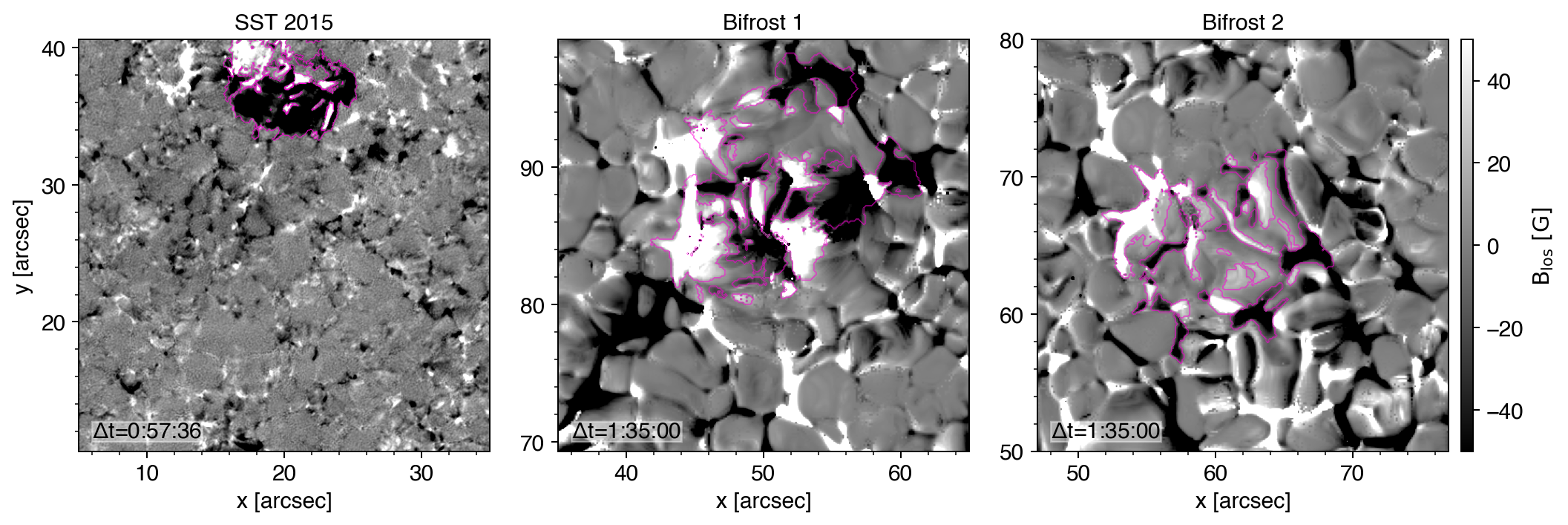}}
	\caption{Snapshot from the animation of the \ion{Fe}{1}~6173~\AA\/ LOS magnetic field maps showing emerging magnetic fields in the photosphere. The left panel shows the emerging ER in SST15. The middle and right panels display a larger and a smaller emerging field in the Bifrost simulation at 100~km horizontal resolution, respectively. For easier visual comparison, the magnetic field maps are scaled to the same values ($\pm50$~G), and the FOVs expand to the same extent. The purple contours enclose footpoints of the emerging bipoles. An animation of this figure is available and runs from $\Delta t = 0:00:00$ to $\Delta t = 2:32:10$. The real-time duration of the animation is 11 s. A higher-resolution animation has been made available on Zenodo at \url{https://zenodo.org/records/15847673}.}
	\label{fig:fig_me_example}
\end{figure*}

\begin{figure*}[!t]
	\centering
	\resizebox{1\hsize}{!}{\includegraphics[width=1.0\textwidth]{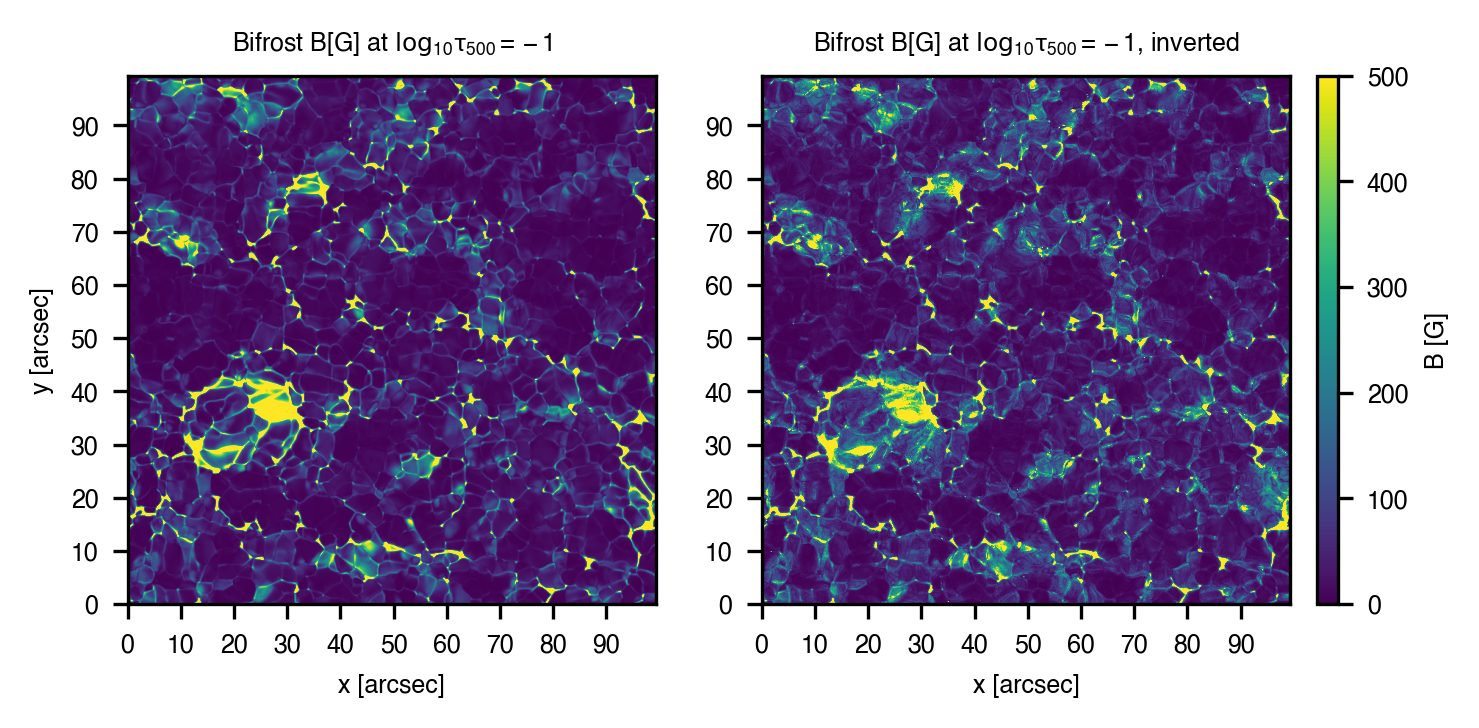}}
	\caption{Maps of the magnetic field strengths at $\log_{10}\tau_{500}=-1$ in a Bifrost snapshot extracted from the {\tt nw072100} model at 100~km horizontal spatial resolution. The left panel shows the original strengths from the simulation, while the right panel shows the strengths derived through SIR inversions of the Bifrost \ion{Fe}{1} 6173 spectral profiles.}
	\label{fig:sir_sim_vs_sim}
\end{figure*}

IRIS data obtained on 2016 August 3, from 07:29:15~UT to 11:59:53~UT, in coordination with SST16 observations, are available but were not used in this work.

All IRIS data sets were corrected for dark current, flat field, geometric distortion, and scattered light, as explained by \cite{2018SoPh..293..149W}. The calibration pipeline included wavelength calibrations and subtraction of the background light leakage in the FUV data. 

\subsection{Bifrost models}

The simulations described in this paper were generated with the Bifrost code \citep{Gudiksen_etal2011}, which solves the equations of radiative MHD on a Cartesian grid in a computational box. The models discussed span the upper convection zone to the corona in a box of physical dimensions $72\times 72\times60$~Mm$^3$ extending $8.5$~Mm below the photosphere to some $52$~Mm above. The computational box is split into a grid consisting of $720\times 720\times 1120$ zones, giving a resolution of 100~km in the horizontal ($xy$) planes. The vertical resolution is variable, with 20~km resolution in the photosphere, chromosphere, and TR, but increasing to an order of 100~km in the larger-scale-height portions of the convection zone and corona. Higher-horizontal-resolution (50~km) runs of the same simulations described here have been computed for limited time spans to examine the effects of resolution on our results. Horizontal boundaries are periodic, while on the lower convection zone boundary, we specify the entropy and the horizontal component of the magnetic field in regions of inflowing gas. Characteristic boundary equations are solved at the upper boundary, as described in \citet{Gudiksen_etal2011}. 

There are two models included in this study, which vary in terms of the magnetic field injected at the lower boundary and spatial resolution in the horizontal planes. In both models, an input entropy commensurate with an effective temperature close to the solar value of $T_{\rm eff}\approx 5800$~K is injected in regions of inflow. In the QS-like model ({\tt qs072050} with 50~km resolution), a constant input magnetic field of $B_{y0}=336.3$~G is continually injected in regions of inflow, while in regions of outflow, the field is allowed to exit the computational domain. The field is injected in order to attempt to achieve an average photospheric magnetic field strength not too different from what is implied by observations, taking into account that any small-scale dynamo is very slow acting in models with fairly coarse resolution, such as the models considered here.

The other model ({\tt nw072100}) uses the same injection method but has a stronger input magnetic field for an extended time span. The model run has an initial 95 minutes period of relatively weak ($B_{y0}=100$~G) magnetic  field injection in inflow regions at the bottom boundary ($z=8.5$~Mm below the photosphere). The injected field is first ramped up to $B_{y0}=1000$~G for 70 minutes, then further increased to $B_{y0}=2000$~G for another 150 minutes, before being reduced to $B_{y0}=336.3$~G for the remainder of the run. This model has been run for more than 10~hr of solar time. In Figure~\ref{fig:nw_overview_464} we show the state of the photosphere and chromosphere some 4 hr after the stronger field was injected in the {\tt nw072100} Bifrost model. Inspection shows that some of this flux has reached or pierced the photosphere and is perturbing photospheric velocities, as well as emission in the wing of the \FeIHMIline\ line and in the \MgIIk\ line. The green rectangles in the figure show two regions of bipolar flux emergence (Bifrost~1 and Bifrost~2): we will concentrate our attention on these in the following.

The optically thick synthetic observables discussed here--e.g. the \FeIHMIline, \CaIIIR\, \CaIIK, and \MgIIk\ lines including Stokes profiles--have been computed using the RH~1.5D code \citep{PereiraUitenbroek2015}. The RH code computes the equations of radiative transfer for each vertical column in the model atmospheres (i.e., all vertical columns in every snapshot considered) independently. \FeIHMIline\ was computed assuming local thermodynamic equilibrium (LTE), while the three chromospheric lines are all computed using non-LTE. Optically thin lines, such as \SiIV, are computed using atomic parameters from the Chianti atomic database package \citep{Chianti10} so that the emergent intensity
$$I_\nu = \frac{h\nu}{4\pi}\int \phi_\nu(u,T)n_{\rm e}n_{\rm H}f(n_{\rm e},T)dz$$
\noindent
where  $\phi_\nu$ is the (Gaussian) emission-line profile, $n_{\rm e}$ and $n_{\rm H}$ are the electron and total hydrogen densities, and $f(n_{\rm e},T)$ contains atomic parameters including the abundance, excitation rates, and ionization state.

\section{Identification and tracking of magnetic features}
\label{tracking}

We detected and tracked all the magnetic elements in the SST15 and SST16 observations, to study their spatiotemporal evolution, and isolated those flux patches that belong to the largest flux emerging region detected in the SST15 data set. This ER was used for comparison with two bipoles emerging in Bifrost snapshots extracted from the {\tt nw072100} model. All three investigated flux emerging systems appeared at the photospheric level as clusters of multiple magnetic elements.

To identify and track flux features in the photosphere, we used line-of-sight (LOS) magnetic field maps calculated by applying the Milne--Eddington (ME) method on the SST and {\tt nw072100} Bifrost model (see Section~\ref{inversions_ME_sst_bifrost} for details). These maps represent the vertical component of the emerging fields, i.e., the footpoints of the emerging magnetic loops. In total, we analyzed 115 snapshots from the Bifrost {\tt nw072100} model at the horizontal spatial resolution of 100~km and a cadence of 50 s. For the SST15 observations, we analyzed 292 images at a cadence of about 24 s. The number of images used to track the magnetic elements in the SST16 data set was 325 at a cadence of 20 s. Stokes $Q$ and $U$ were employed to analyze the spatiotemporal evolution of the horizontal fields, corresponding to the tops of the emerging loops. We applied a $3\sigma$ threshold on the LOS magnetic field maps, discarding pixels where the signal fell below this value. The noise level, $\sigma$, was taken from the noisier data set (SST15) and is determined to be 5~G by computing the standard deviation of the flux density in a region without clear solar signals. The identification method for automatically identifying newly appeared magnetic elements was the downhill method, while the tracking of the identified elements from frame to frame was performed using the YAFTA\footnote{\url{https://solarmuri.ssl.berkeley.edu/~welsch/public/software/YAFTA/}} code \citep{2003ApJ...588..620W}. YAFTA assigns a unique label to each magnetic feature, which allows us to trace back the history of the elements detected in the LOS magnetic field maps. This enables us to calculate various characteristics, including their modes of appearance and disappearance, interactions, unsigned magnetic flux contents, horizontal velocities, sizes, and lifetimes. While the tracking process is automated, we manually verified the clusters to ensure accuracy. We considered only magnetic elements with a minimum size of 5 pixels. Additionally, we employed the linear polarization maps to identify the locations where new magnetic loops emerged. To classify an element as part of a cluster of magnetic features, it must appear in situ within a group of mixed-polarity patches in a relatively small area and move outward along the cluster's magnetic axis. 

For each detected magnetic element, we examined its interactions with other magnetic elements on the surface, to investigate the dynamics of the photospheric footpoints of the detected loops. In particular, we determined the numbers of the following types of surface interactions experienced by magnetic elements during their lifetimes:

\begin{enumerate}
    \item Merging, or the coalescence of two or more elements of the same polarity into a larger structure. 

    \item Fragmentation, the opposite of merging, which occurs when an element splits into two or more smaller features.

    \item Cancellation, or the disappearance of a magnetic element in the vicinity of an opposite-polarity feature. Through cancellations, magnetic flux is either partially or completely removed from the photosphere.
\end{enumerate}

Magnetic elements that fragment from the footpoints are classified as cluster members. Flux patches that merge with the footpoints but appear far from the flux emergence centers are excluded from the clusters. 

\section{Inversions of SST and Bifrost Data}
\label{inversions_sst}

\subsection{ME inversions}
\label{inversions_ME_sst_bifrost}

To compare the spatial and temporal properties of the magnetic fields detected in the SST sequences and the simulation {\tt nw072100} snapshots, we performed a fast and robust pixel-to-pixel ME inversion of the \ion{Fe}{1}~6173~\AA\/ line using the PyMilne\footnote{\url{https://github.com/jaimedelacruz/pyMilne}} code \citep{2019A&A...631A.153D}. The ME method assumes that atmospheric parameters remain constant with height, while the source function changes linearly with optical depth. 

The simulations were not convolved with the SST spatial point-spread function (PSF), since the SST data reduction process already includes spatial deconvolution. We also did not regrid the synthetic maps to match the pixel size of the SST detector, because SST has a spatial resolution of $~83$~km ( apixel size of $0\farcs057$), which is smaller than the pixel size of the Bifrost simulations ($100$~km). 

However, we spectrally degraded each synthetic profile by convolving it with the spectral SST PSF. The convolved profiles were then interpolated to match the wavelength sampling of the SST measurements. We selected only the wavelengths present in SST15 for the inversions. 

An example of the ME inversions is presented in Figure~\ref{fig:fig_me_example}, which shows three regions of emerging flux. The left panel reveals the emergence of new flux in the photosphere as observed by SST, while the middle and the right panels depict, respectively, one larger and one smaller event of emerging flux in the Bifrost {\tt nw072100} model. The violet contours enclose the footpoints, including only pixels above 15~G, a threshold selected based on a $3\sigma$ noise level derived from the noisier SST15 data set.

\subsection{SIR inversions}

A more precise comparison of the magnetic and thermodynamic properties of the solar atmosphere, derived from the Bifrost simulations and SST observations, is carried out by inverting the data using the SIR\footnote{\url{https://github.com/BasilioRuiz/SIR-code}} code \citep{1992ApJ...398..375R}. For this purpose, we inverted the \ion{Fe}{1} 6173 Stokes parameters from one Bifrost snapshot and one SST frame in which the emerging fields are well developed. SIR is based on the Levenberg-Marquardt algorithm, which minimizes the difference between measured and computed synthetic Stokes profiles using response functions. It solves the radiative transfer equation under the assumption of local thermodynamic equilibrium (LTE) and returns the temperature, velocity, magnetic field strength, inclination, and azimuthal angles along the LOS.

We used a simple one-component model atmosphere and two inversion cycles. In the first cycle, we inverted the \ion{Fe}{1} line with a single node for the magnetic field strength, inclination, azimuth, and LOS velocity, while using two nodes for temperature. We set the magnetic filling factor to 1. In the second cycle, we increased the number of nodes for the magnetic field strength and LOS velocity to two, in order to account for the asymmetries between the blue and red lobes of Stokes $V$. In this cycle, the number of nodes for temperature was three. 
Having multiple nodes for temperature, magnetic field strength, and LOS velocity allowed us to achieve reasonable fits to some of the more complex profiles found in the regions where flux was emerging. Without this setup, we would struggle to estimate the LOS velocities accurately, and in some pixels, the magnetic field strength would be significantly underestimated.
For the initial guess of the model atmosphere, we used either the Harvard Smithsonian Reference Atmosphere \citep[HSRA;][]{1971SoPh...18..347G} or the FALC model \citep{1993ApJ...406..319F}, depending on which provided better results, as indicated by the $\chi^{2}$ values calculated by the SIR code.

\begin{figure}[!t]
	\centering
	\resizebox{1\hsize}{!}{\includegraphics[width=1.0\textwidth]{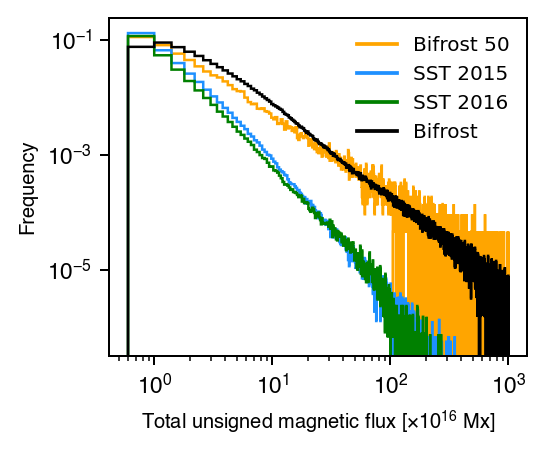}}
	\caption{The unsigned magnetic flux distributions for the SST15 (blue), and SST16 (green) data sets. The corresponding distributions derived from the Bifrost snapshots are presented with the black (100~km resolution) and orange (50~km resolution) lines. The bin sizes are
	$4 \times 10^{15}$~Mx.}
	\label{fig:fl_fov}
\end{figure}

\begin{figure}[!t]
	\centering
	\resizebox{1\hsize}{!}{\includegraphics[width=1.0\textwidth]{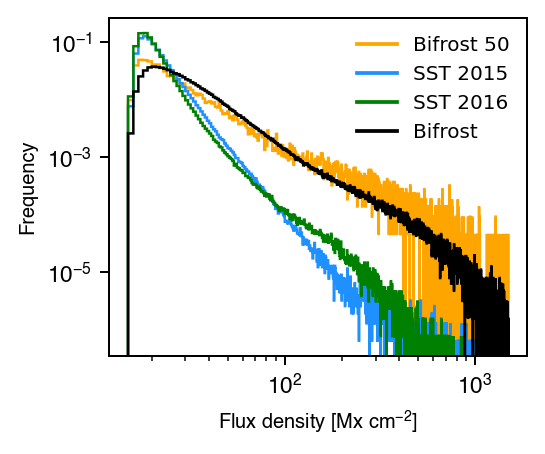}}
	\caption{The mean magnetic flux density distributions for the SST15 (blue) and SST16 (green) observations and the Bifrost simulations (the black and orange lines refer to the 100 km and 50 km resolution variants, respectively). The bin sizes are $1$~Mx~cm$^{-2}$.}
	\label{fig:fld_fov}
\end{figure}

\begin{figure}[!t]
	\centering
	\resizebox{1\hsize}{!}{\includegraphics[width=1.0\textwidth]{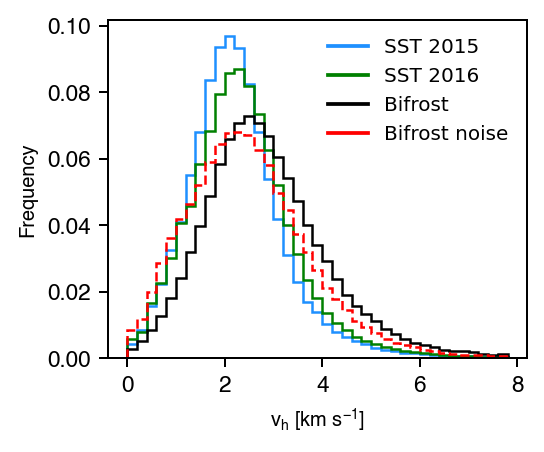}}
	\caption{The mean horizontal velocities of the magnetic features detected in the SST15 (blue), SST16 (green), and Bifrost (black) data sets. The red dashed line shows the mean horizontal velocity distribution for the Bifrost data after a Gaussian noise with a sigma of 5~Mx~cm$^{-2}$ is added to the Bifrost images. The bin sizes are 0.2~km~s$^{-1}$}
	\label{fig:vh_fov}
\end{figure}

To confirm that SIR is capable of inverting the Bifrost \ion{Fe}{1} 6173 spectral profiles, we inverted one snapshot without any degradation, using only those wavelengths available in the SST15 measurements. This approach simplified the shape and complexity of the spectral lines to some extent. Nevertheless, SIR managed to derive the model atmosphere parameters comparable to the original Bifrost values. In Figure~\ref{fig:sir_sim_vs_sim}, we present maps of the magnetic field strengths at $\log_{10}\tau_{500}=-1$ for both the original Bifrost {\tt nw072100} model atmosphere (left panel) and the model atmosphere derived from SIR inversions (right panel). The Pearson linear correlation coefficient for these two maps is 0.93.

\section{Results}
\label{results}

\subsection{Tracking results}

The first step in understanding the QS magnetic fields is analyzing the properties of the patches identified in the magnetograms. The key parameters to consider include their total unsigned fluxes, flux densities, sizes, and lifetimes. Additionally, we calculated the average horizontal velocities and the total number of surface interactions for each detected feature. Finally, we estimated the flux appearance and disappearance rates and the total number of canceling flux features.

We analyzed the magnetic features in four datasets, two from the SST and two from the Bifrost simulation sequences of QS regions. Considering the entire available FOVs, we found more than 200,000 unique magnetic features in three datasets, while in the 50~km Bifrost snapshots, we detected 16,241. Among the observations, the SST15 data had the highest number of detected features (513,486), while the Bifrost 100~km snapshots had 200,294. If we differentiate between features and patches, where a feature is recognized as a series of individual magnetic patches in consecutive time steps, then the numbers of detected patches vary significantly, ranging from 21,368 (Bifrost, 50~km) to 623,860 (Bifrost, 100~km) and 1,534,553 (SST15). The SST16 data set contains 1,276,057 patches.

\subsubsection{Flux content}

The flux distributions of the detected patches in the SST data are shown in Figure \ref{fig:fl_fov}. These patches span nearly three decades of flux, ranging from the lower detection limit of approximately $1 \times 10^{15}$~Mx to about $6 \times 10^{18}$~Mx. The Bifrost data (the black and orange solid lines) indicate the presence of the strongest detected flux patches, with a maximum flux content of around $1 \times 10^{19}$~Mx in the 50~km snapshots and $3 \times 10^{19}$~Mx in the 100~km snapshots. The decrease in flux at the lower end of the distributions is twofold: (1) due to sensitivity limitations of the observations; and (2) due to thresholding (for both the SST and Bifrost data sets). Since the distributions include all the frames and the entire FOVs, the rate at which weaker fields are generated in Bifrost simulations may also be affected by varying magnetic field injections at the bottom boundary. Conversely, the drop-off at the high end of all four flux distributions represents a reduction in the strongest flux concentrations. On average, the total flux of the magnetic patches is about $2 \times 10^{16}$~Mx in the SST observations and 1 order of magnitude larger in the Bifrost models. 

\begin{figure*}[!t]
    \begin{subfigure}{}
    \includegraphics[width=0.45\linewidth]{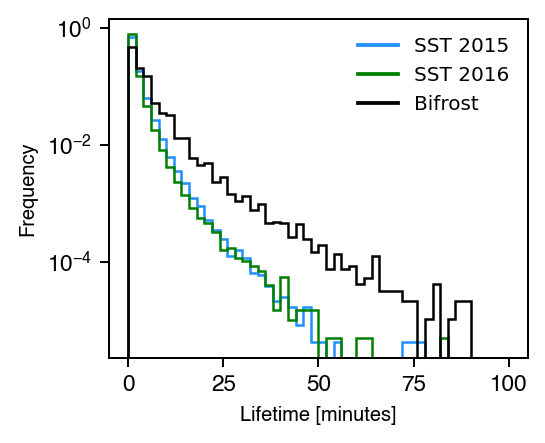} 
    \label{fig:sub_lt}
    \end{subfigure}
    \begin{subfigure}{}
    \includegraphics[width=0.45\linewidth]{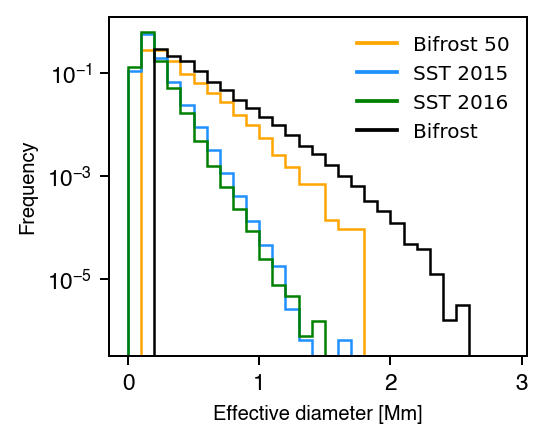}
    \label{fig:sub_size}
    \end{subfigure}
	\caption{Lifetime and effective diameter distributions of flux features detected in the SST15, SST16, and Bifrost data sets. The distributions are shown with blue, green, black (100~km), and orange (50~km) lines, respectively. The bin sizes are 2 minutes for lifetimes and 0.1~Mm for effective diameters.}
	\label{fig:lt_size_fov}
\end{figure*}

The SST magnetic field maps display a combination of weak IN and stronger NW fields, whereas the Bifrost data predominantly feature strong, NW-like fields. However, the 50~km Bifrost run contains somewhat weaker fields compared to the 100~km variant. Elevated flux content is typically associated with stronger, kG, fields, which are commonly found in NW regions of the QS.

The detection of stronger magnetic fields in the Bifrost simulations compared to observations is further supported by Figure \ref{fig:fld_fov}, which displays the mean unsigned flux density distributions for the detected magnetic elements. The average values for the datasets are as follows: 24~Mx~cm$^{-2}$ (SST15), 23~Mx~cm$^{-2}$ (SST16), 64~Mx~cm$^{-2}$ (Bifrost), and 78~Mx~cm$^{-2}$ (Bifrost, 50~km). The cutoff at 15~Mx~cm$^{-2}$ results from the $3\sigma$ threshold applied to identify magnetic patches. The maximum mean flux densities recorded are 1340~Mx~cm$^{-2}$, 859~Mx~cm$^{-2}$, 3238~Mx~cm$^{-2}$, and 3425~Mx~cm$^{-2}$ for the SST15, SST16, Bifrost, and Bifrost 50~km snapshots, respectively.

The mean unsigned flux density in the Bifrost snapshots, considering all the pixels within the FOV, is 32.6~Mx~cm$^{-2}$ at 100~km spatial resolution and 28~Mx~cm$^{-2}$ at 50~km resolution. The observations show lower values, 11.8~Mx~cm$^{-2}$ in the SST15 and only 9.3~Mx~cm$^{-2}$ in the SST16 data sets. Again, these values suggest that the Bifrost simulations do not represent the average state of QS, but mainly represent NW fields.

The unsigned longitudinal magnetic flux density has been used historically to estimate the magnetization of the QS and to compare observations with numerical models. Unfortunately, there is a large scatter among the values reported in the literature. This results from several limitations: atmospheric seeing, the noise of the observations, the sensitivities of the spectral lines used to record the measurements, diagnostic techniques, and systematic instrumental and data processing errors. In addition, the flux density estimates also depend on the filling factor of the observations, which is different from one. Because of this, the longitudinal flux densities derived from observations and Zeeman-sensitive spectral lines should be considered as a lower limits, and are reported in the literature as ranging from 1-2 to about 30~Mx~cm$^{2}$ \citep[e.g.,][]{1994A&A...286..626K, 1995SoPh..160..277W, 2003ApJ...582L..55D, 2008A&A...480..265B, 2009A&A...502..969B, 2009ApJ...690..279J, 2012ApJ...746..182O}. Observations based on the Hanle effect lead to higher flux density values beyond 60~Mx~cm$^{2}$ \citep{2004Natur.430..326T}. 

Contrary to the observations, the filling factor in numerical simulations is always equal to one, but the mean flux density also depends on the input fields and the physics implemented to further drive the generation of magnetic structures. The (horizontal) resolutions of the Bifrost models are relatively coarse, at $100$ and $50$~km respectively, hindering these models from accurately describing dynamics at scales smaller than this. It is not entirely clear what the consequences of the coarse resolution are: models with much better resolution \citep[e.g.,][]{2014ApJ...789..132R,2016A&A...593A..93D} indicate that an average vertical $B_z$ consistent with the observations is of order $60$~G, much larger than the models presented here. On the other hand, it is not clear how much impact photospheric fields on such small scales have on the averages at coarser resolutions, nor on the dynamics and energetics of the chromosphere or corona. The extension of modeling to both higher resolution and varying average field strengths will be investigated in future studies.

\subsubsection{Flux appearance and disappearance rates}

In addition to calculating the mean flux density, the flux balance on the solar surface can also be evaluated by examining the rates at which magnetic flux appears and disappears, specifically through the processes that bring flux to and remove it from the photosphere. Newly generated magnetic structures emerge in situ at the surface as magnetic bipoles. Additionally, many elements appear as unipolar patches in observations, due to instrumental or feature tracking limitations \citep[see, e.g.,][]{2002ApJ...569..474D, 2016ApJ...820...35G, 2022ApJ...925..188G}. Consequently, the rates of flux appearance and disappearance may be overestimated, unless a careful analysis of the history and interactions of magnetic fields is conducted, as explained in \cite{2016ApJ...820...35G} and \cite{2022ApJ...925..188G}. In our Bifrost simulations, magnetic elements appear in situ due to magnetic buoyancy, convective upwelling, or local dynamo action. Flux disappears from the solar surface through two main processes: fading (in situ disappearance) and cancellation (the total or partial elimination of opposite-polarity magnetic features when they come into close proximity).

To calculate the flux rates, we consider the flux content of magnetic elements at the moment they first become visible in the photosphere (in situ appearance) and when they completely disappear. In this context, the disappearance rate encompasses both in situ fading and total cancellation of magnetic elements. Based on our estimations, we find that magnetic flux appears in the photosphere at the following rates: 303 Mx cm$^{-2}$ day$^{-1}$ (SST15), 327 Mx cm$^{-2}$ day$^{-1}$ (SST16), and 175 Mx cm$^{-2}$ day$^{-1}$ (Bifrost). We can assume that, on average, the magnetic elements in the QS increase in flux by a factor of three over their lifetimes \citep{2016ApJ...820...35G}. Taking into account this intrinsic flux growth of magnetic features, our estimated flux appearance rates are within the limits reported in the literature. The disappearance rates for the SST15 and SST16 data sets and the Bifrost simulations are 292~Mx~cm$^{-2}$~day$^{-1}$, 316~Mx~cm$^{-2}$~day$^{-1}$, and 160~Mx~cm$^{-2}$~day$^{-1}$, respectively. These rates are comparable to their corresponding appearance rates, indicating that the magnetic flux in all data sets is in balance.

\subsubsection{Horizontal velocity}

On average, the Bifrost magnetic patches exhibit dynamic properties similar to the fields detected in the SST data, even though they are associated with stronger fields. The mean horizontal velocities derived from the SST15, SST16, and Bifrost datasets are 2.1, 2.3, and 2.8~km~s$^{-1}$, respectively (Figure \ref{fig:vh_fov}). These values are calculated by measuring the distances traveled by the flux-weighted centers of the detected magnetic features between the consecutive frames. It is important to note that the observed displacements of magnetic patches are not only determined by their intrinsic evolution, but are also driven by supergranular advection, granular motions, and random jitter caused by surface interactions between magnetic patches \citep{2018ApJ...854..118A}. 

The horizontal velocities in the Bifrost data are slightly more spread and shifted toward higher values, having a maximum velocity of about 29~km~s$^{-1}$, compared to 26~km~s$^{-1}$ in the SST16 and 20~km~s$^{-1}$ in the SST15 datasets. 
One of the factors that may influence the velocities is noise. The downhill method, used in this work, is effective at identifying individual patches from clusters of magnetic flux, but noise fluctuations can create small local maxima that the downhill method might incorrectly identify as new structures \citep{2007ApJ...666..576D}. Some other identification methods may reduce these kinds of errors but introduce others \citep{2007ApJ...666..576D, 2014ApJ...797...49G, 2016ApJ...820...35G}. After adding a Gaussian noise with a sigma of 5~Mx~cm$^{-2}$ to the Bifrost images, the mean horizontal velocity shifted slightly to lower values, having a mean value of about 2.4~km~s$^{-1}$ (the red dashed line in Figure~\ref{fig:vh_fov}). 

\subsubsection{Lifetime and size}

The left panel of Figure~\ref{fig:lt_size_fov} shows the mean lifetimes of the detected flux features, and the values are 1.8 (SST15), 1.5 (SST16), and 3.9 (Bifrost) minutes. The respective cadences determine the shortest lifetimes, and they are 20~s for the SST16 data set, 24~s for SST15, and 50~s for the Bifrost data. Since higher cadences allow for better resolving of the temporal evolution of individual magnetic features, the mean lifetime is expected to decrease as cadence increases. It is also expected that larger features (such as in Bifrost) will have longer lifetimes on average, because they are less susceptible to being dispersed by convective motions.

The right panel of Figure~\ref{fig:lt_size_fov} shows the mean effective size\footnote{The effective diameter represents the diameter of a circular structure with the same area as the specified magnetic patch.} distribution for the three data sets (SST15, SST16, and the Bifrost data at 100~km spatial resolution) and, in addition, for the 50~km Bifrost snapshots. On average, the largest flux patches are found in the Bifrost simulations at 100~km resolution (0.41~Mm), followed by the 50~km Bifrost model (0.28~Mm). The SST15 and SST16 data sets have mean effective diameters of 0.13~Mm and 0.12~Mm, respectively. The differences between the sizes of the 100~km and 50~km patch sizes as well as the differences between the Bifrost patches and the observed ones suggest that the coarse resolution of the models may play an important role. Overall, magnetic structures appear to be almost a factor of 2 larger than those seen in the SST data cubes. This will be the subject of future studies.
\vspace*{1em}

\begin{figure*}[!t]
    \begin{subfigure}{}
    \includegraphics[width=0.45\linewidth]{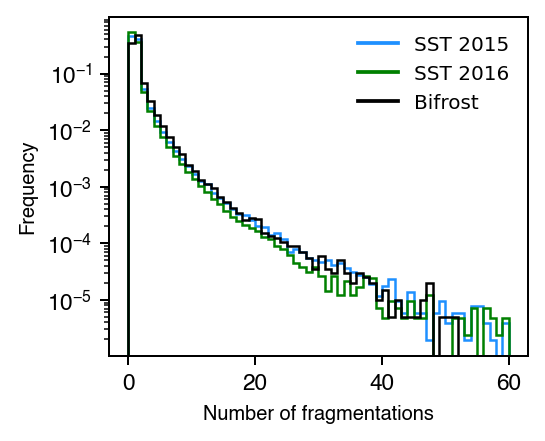} 
    \label{fig:sub_frag}
    \end{subfigure}
    \begin{subfigure}{}
    \includegraphics[width=0.45\linewidth]{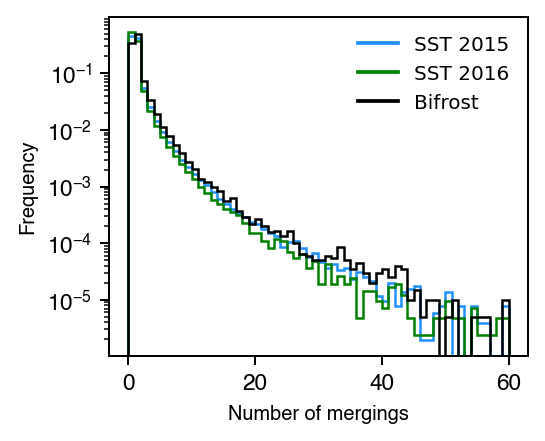}
    \label{fig:sub_merg}
    \end{subfigure}
\caption{Distribution of the total number of fragmentation (left) and merging events (right) of magnetic features in the QS. They are shown separately for SST15 (blue), SST 2016 (green), and Bifrost (black curve) features. The used bin sizes are 1.}
\label{fig:frag_merg_fov}
\end{figure*}

\subsubsection{Number of surface interactions}

During their lifetimes, magnetic elements may fragment and merge with other flux features. The number of fragmentation (left panel) and merging events (right panel) that individual flux features experience over their lifetimes is displayed in Figure \ref{fig:frag_merg_fov}. As can be seen, all the data sets have practically the same number of interactions per magnetic feature, where most of them either do not interact with other features or have only one or two interactions before they disappear. 
 
The main reason why that we find such a limited number of interactions is the downhill method, rather than granular dynamics. The downhill method focuses on following local flux maxima and effectively ignores surface interactions, unless the features clearly fragment or completely merge with others of the same polarity. This method thus sets a high threshold for the detection of an interaction.

\subsection{Magnetic bipoles in SST and Bifrost data sets}

\begin{figure}[!t]
	\centering
	\resizebox{1\hsize}{!}{\includegraphics[width=1.0\textwidth]{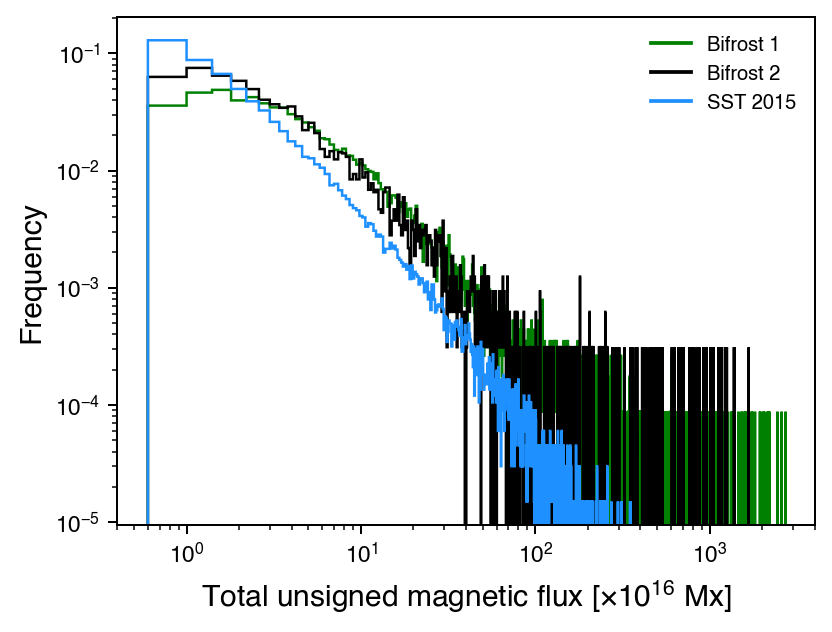}}
	\caption{The unsigned magnetic flux distributions for the patches forming two bipoles identified in the Bifrost snapshots (green and black lines) and the strongest bipoles observed in the SST15 (blue) data set. The bin sizes are $4 \times 10^{15}$~Mx.}
	\label{fig:fl_fp}
\end{figure}

\begin{figure}[!t]
	\centering
	\resizebox{1\hsize}{!}{\includegraphics[width=1.0\textwidth]{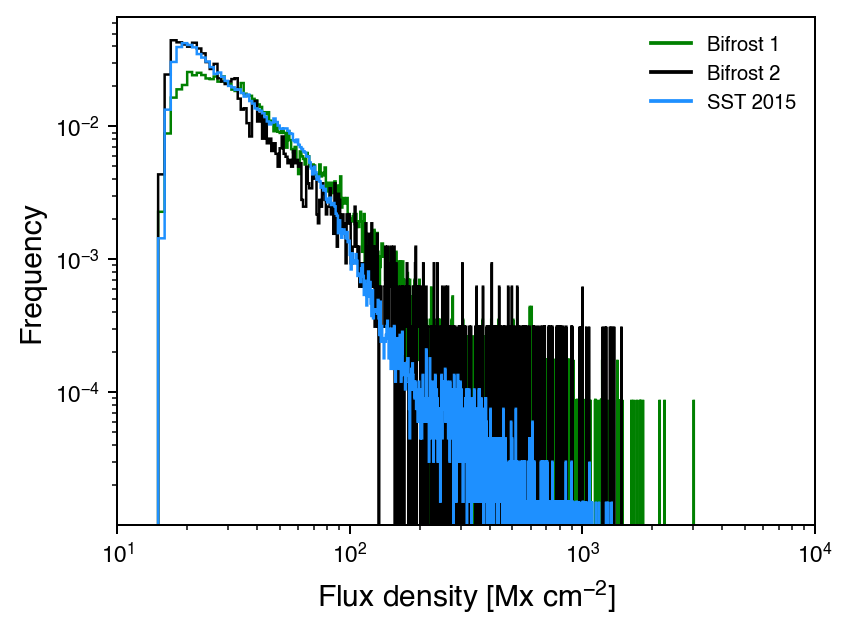}}
	\caption{The same as Figure~\ref{fig:fl_fp} but for the mean magnetic flux density. The bin sizes are $1$~Mx~cm$^{-2}$.}
	\label{fig:fld_fp}
\end{figure}

The main reason for developing the Bifrost models described in this work is to study how the chromosphere and corona respond to the magnetic fields that emerge from below. Therefore, we will now focus on comparing several magnetic bipoles identified in the Bifrost {\tt nw072100}  model and SST data sets.  

The flux distributions of the magnetic patches that form the largest emerging bipoles in the SST data, along with two Bifrost bipoles, are shown in Figure \ref{fig:fl_fp}. The flux content of the individual SST footpoint patches ranges from approximately $1 \times 10^{15}$~Mx to about $3.5 \times 10^{18}$~Mx. In contrast, the Bifrost bipoles (represented by the green and black solid lines) exhibit larger flux patches, with a maximum flux content of around $3 \times 10^{19}$~Mx. Despite this, all three bipoles contain a significant number of medium strong flux patches (greater than $10^{17}$~Mx and smaller than $10^{18}$~Mx) and large flux patches ($>10^{18}$~Mx). As a result, their unsigned flux distributions are more similar to each other than those obtained when considering the entire FOVs. The average total unsigned fluxes for the larger and weaker Bifrost bipoles are $35 \times 10^{16}$~Mx and $27 \times 10^{16}$~Mx, respectively. In contrast, the footpoint patches measured in SST15 exhibit lower average flux contents of $7 \times 10^{16}$~Mx. 

Figure \ref{fig:fld_fp} demonstrates that the flux patches in the three bipoles possess flux densities of the same order. The mean flux density values for SST15, Bifrost~1 (larger bipole), and Bifrost~2 (smaller bipole) are 47~Mx~cm$^{-2}$, 80~Mx~cm$^{-2}$, and 64~Mx~cm$^{-2}$, respectively. The maximum flux density distributions for each bipole are as follows: 1340~Mx~cm$^{-2}$ for SST15, 3020~Mx~cm$^{-2}$ for Bifrost~1, and 1475~Mx~cm$^{-2}$ for Bifrost~2.

\begin{figure}[!t]
	\centering
	\resizebox{1\hsize}{!}{\includegraphics[width=1.0\textwidth]{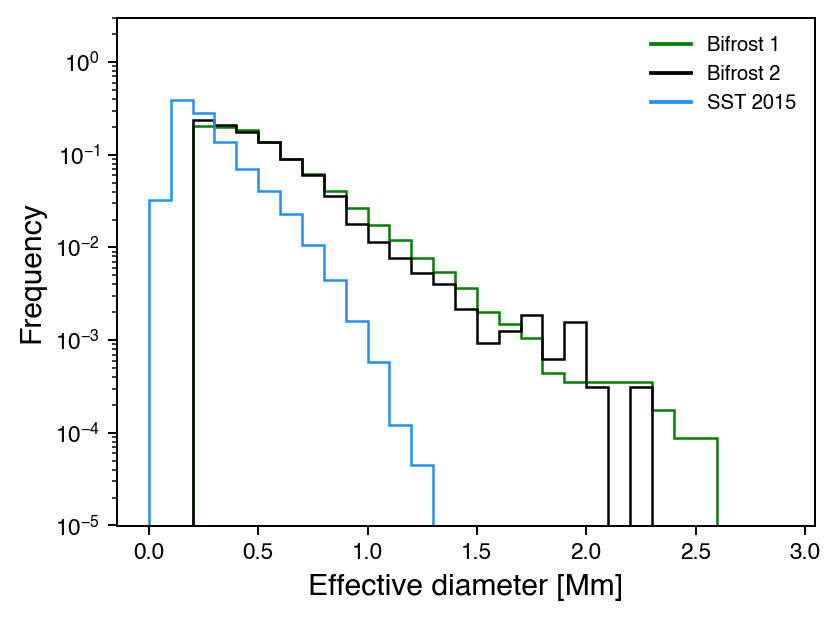}}
	\caption{The same as Figure~\ref{fig:fl_fp} but for the effective diameter of the detected magnetic elements forming bipolar footpoints. The bin sizes are $0.05$~Mm.}
	\label{fig:size_fp}
\end{figure}

The distributions of the mean effective sizes are shown in Figure~\ref{fig:size_fp}. On average, the Bifrost bipoles are twice as large as the magnetic patches observed in the SST15 bipole. The largest patches were found in Bifrost~1, measuring 0.47~Mm, and then Bifrost~2 bipole, measuring 0.44~Mm. The emerging fields in the SST15 dataset have mean patch sizes of 0.21~Mm.

\begin{figure}[!t]
	\centering
	\resizebox{1\hsize}{!}{\includegraphics[width=1.0\textwidth]{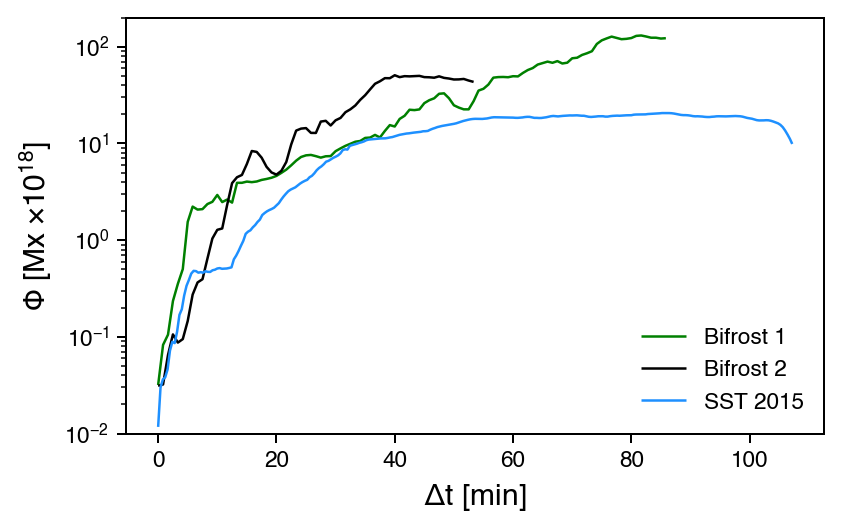}}
	\caption{Instantaneous unsigned magnetic fluxes for the examined SST and Bifrost bipoles.}
	\label{fig:total_fp_dt}
\end{figure}

By identifying the individual magnetic patches that constitute the footpoints, we can calculate the total amount of magnetic flux brought to the surface by the three emerging bipoles, as well as the effective sizes of their entire footpoints. In Figure~\ref{fig:total_fp_dt}, we present the instantaneous flux contained within the footpoints. All three bipoles exhibit a rapid increase in flux during the first approximately 10 minutes of the emerging phase. This increase then slows down and reaches a plateau by the end of the emerging period. The maximum total unsigned fluxes contained by the studied bipoles are approximately $13 \times 10^{19}$~Mx (Bifrost 1), $5 \times 10^{19}$~Mx (Bifrost 2), and $2 \times 10^{19}$~Mx (SST15). These values put all three bipoles in the domain of ERs.

\begin{figure}[!t]
	\centering
	\resizebox{1\hsize}{!}{\includegraphics[width=1.0\textwidth]{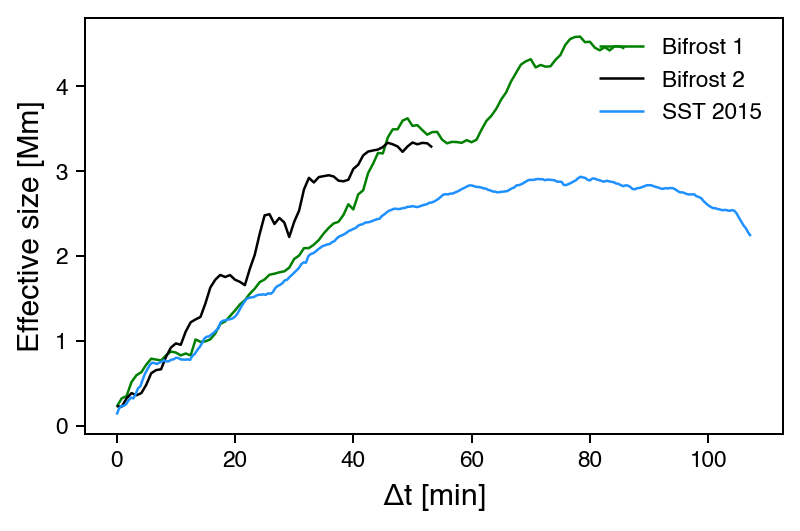}}
	\caption{Effective sizes of the footpoints of the three examined SST and Bifrost bipoles.}
	\label{fig:size_fp_dt}
\end{figure}

Figure~\ref{fig:size_fp_dt} illustrates the growth of the total effective size of the positive- and negative-polarity footpoints over time. As depicted, the largest footpoints are associated with the Bifrost 1 bipole, with an effective size of 4.6~Mm. The other Bifrost bipole is smaller (3.3~Mm), while the footpoints identified in the SST data have an effective size of 2.9~Mm (SST15). The footpoints of the emerging fields have different separation speeds, which are based on the identification of the most distant flux-weighted centers of magnetic patches forming the footpoints. In the Bifrost simulations, the newly emerging footpoints of the larger bipole (Bifrost 1) separate at a mean speed of 2.8~km~s$^{-1}$, while the footpoints within the smaller bipole (Bifrost 2) move away from each other at a speed of 3.2~km~s$^{-1}$. These separation speeds are higher than those observed for the SST15 bipole, which has a speed of 1.2~km~s$^{-1}$. However, the separation speeds of the Bifrost footpoints are comparable to the reported values for QS ERs. These larger bipoles expand at approximately 3.9~km~s$^{-1}$ during their emerging phase and slow down to 0.4~km~s$^{-1}$ afterward \citep{1998Natur.394..152S}. As a reference, smaller IN bipoles on average expand at a speed of about 2~km~s$^{-1}$ \citep{2019LRSP...16....1B}.

\begin{figure*}[!t]
	\centering
	\resizebox{1\hsize}{!}{\includegraphics[width=1.0\textwidth]{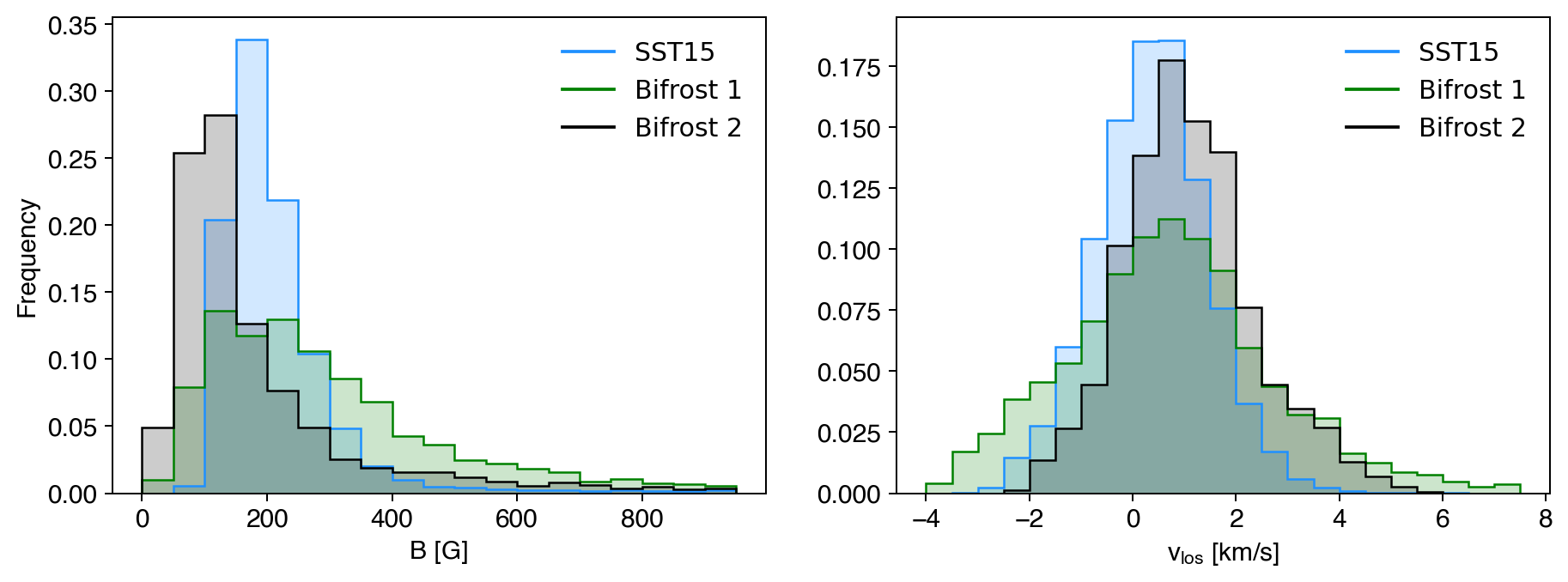}}
	\caption{Magnetic field strength (left panel) and LOS velocity (right panel) distributions for all patches forming the examined Bifrost bipoles (green and black solid lines) and the largest bipoles identified in the SST15 (blue) datasets. The bin sizes are $50$~G and $0.5$~km~s$^{-1}$ for the field strengths and velocities, respectively.}
	\label{fig:b_vlos_sir_fp}
\end{figure*}

\begin{figure*}[!t]
	\centering
	\resizebox{1\hsize}{!}{\includegraphics[width=1.0\textwidth]{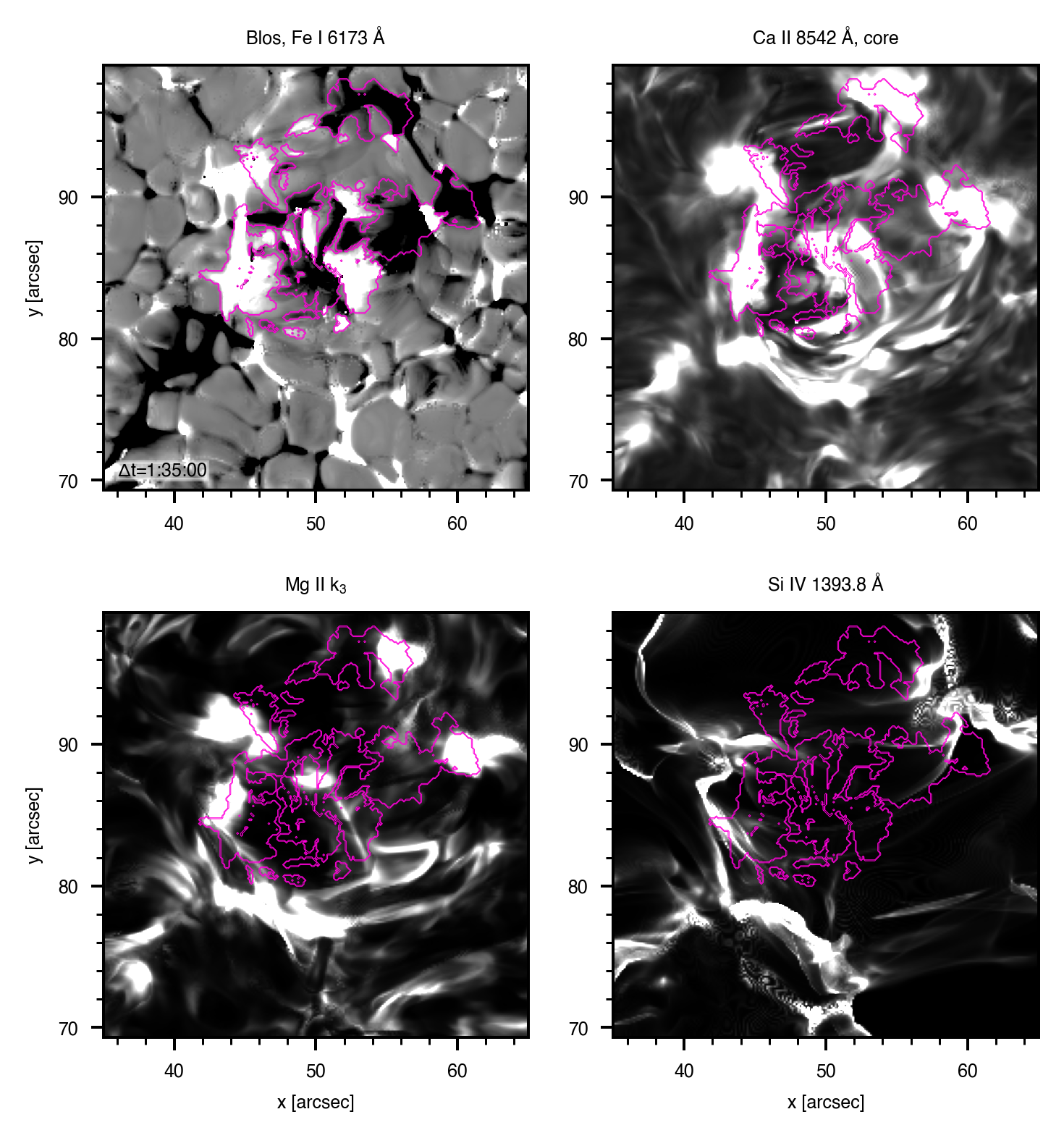}}
	\caption{Snapshots of the signals in the Bifrost \ion{Fe}{1} 6173 \AA\/ LOS photospheric magnetic fields map (upper left panel), the chromospheric \ion{Ca}{2} 8542 \AA\/ intensity in the core of the line (upper right panel), the \ion{Mg}{2}~k$_{3}$ intensity (bottom left panel) and \ion{Si}{4}~1393.8~\AA\/ intensity map (bottom right panel). The violet lines mark the footpoint boundaries of the emerging bipole. The filtergrams display the Bifrost 1 bipole after it emerged and fully expanded. Strong emission is visible at the chromospheric heights and is cospatial with the photospheric footpoints and polarity inversion lines.}
	\label{fig:fe_ca_mg_bifrost1}
\end{figure*}

The distributions of magnetic field strength, derived from the SIR inversions of the Stokes profiles in the \ion{Fe}{1} 6173 line, are illustrated in the left panel of Figure~\ref{fig:b_vlos_sir_fp}. All pixels within the footprints are considered. As can be seen, the SST15 and Bifrost bipoles exhibit similar average field strengths: about 190 G for SST15, 260 G for Bifrost 1, and 150 G for Bifrost 2. Notably, the stronger Bifrost bipole contains more pixels with higher field strengths, ranging from hectogauss to kG values, compared to the other two analyzed bipoles.

\begin{figure*}
	\centering
	\resizebox{0.95\hsize}{!}{\includegraphics[width=1.0\textwidth]{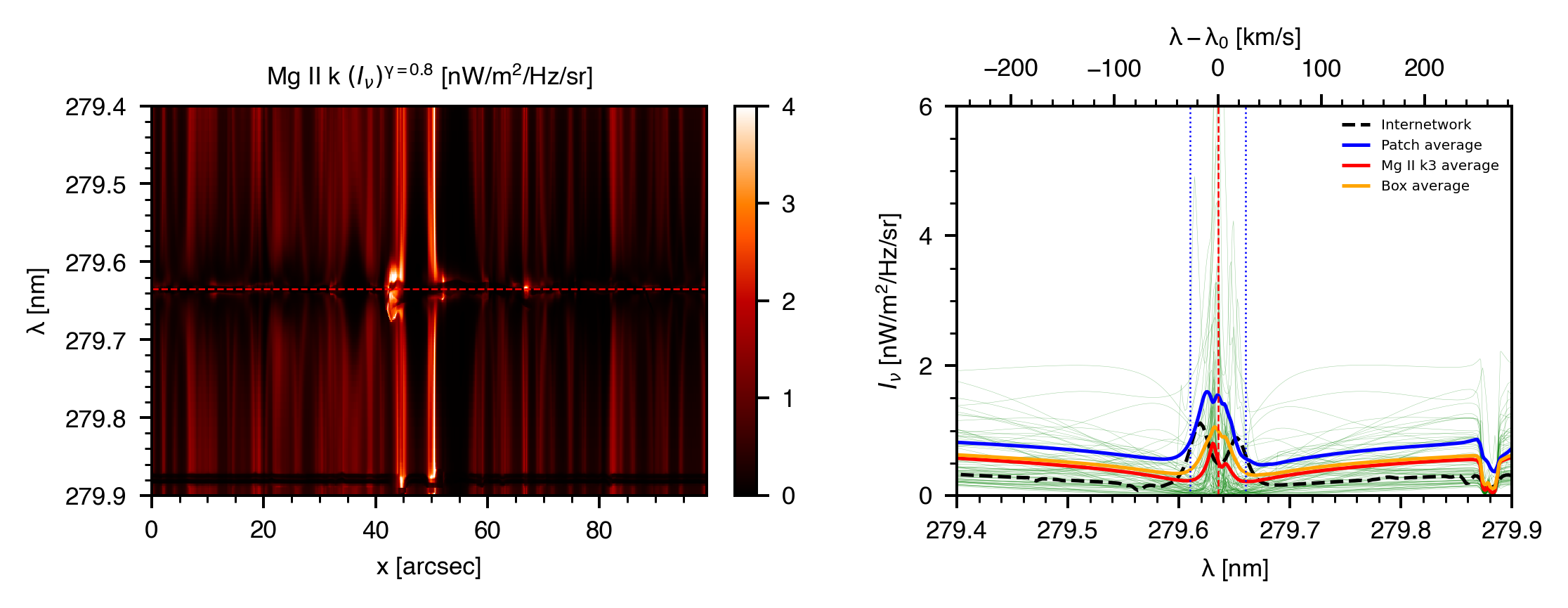}}
	\caption{\MgIIk\ and \MgIItriplet\ line profiles in the Bifrost~1 region and along the dashed green line in Figure~\ref{fig:nw_overview_464} (left panel). Gamma correction ($\gamma$) of 0.8 is used to adjust the overall image brightness and contrast. In the right panel, the red line shows the average \MgIIk\ line profile for the entire computational domain, while the blue line shows the average line profile within the footpoints of the Bifrost~1 bipole. The green lines show the \MgIIk\ profiles in random locations throughout the green box shown in Figure~\ref{fig:nw_overview_464}, and the orange curve represents the average \MgIIk\ profile in the same box. The red dashed lines indicate the central wavelength of \MgIIk, and the dotted blue lines in the right panel show the typical observed k2/k3 core widths (FWHM).}
	\label{fig:mg_profiles_bifrost1}
\end{figure*}

\begin{figure*}
	\centering
	\resizebox{0.95\hsize}{!}{\includegraphics[width=1.0\textwidth]{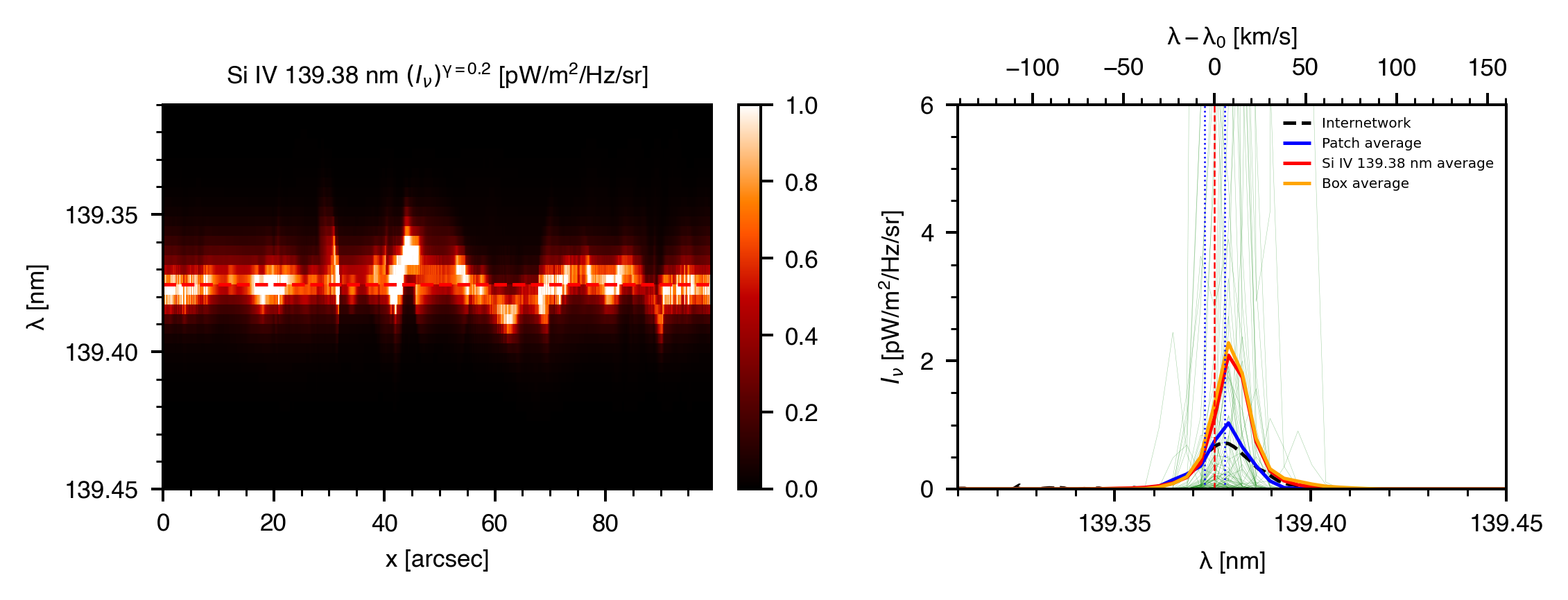}}
	\caption{The same as Figure~\ref{fig:mg_profiles_bifrost1} but for the TR \SiIV\ line. In contrast to the \MgIIk\ figure, the blue dotted lines show the thermal width of the \SiIV\ line rather than any observed width.}
	\label{fig:si_profiles_bifrost1}
\end{figure*}

The right panel presents the LOS velocity distributions for the pixels located within the identified footpoints. The mean LOS velocities ($v_{los}$) are about $0.1$~km~s$^{-1}$ for SST15, $0.5$~km~s$^{-1}$ for Bifrost 1, and $0.8$~km~s$^{-1}$ for Bifrost 2. Even though the mean $v_{los}$ are similar, the distributions for Bifrost show longer tails, particularly for the downflow velocities.

\subsubsection{Bifrost and SST bipoles in the chromosphere}

Figure~\ref{fig:fe_ca_mg_bifrost1} displays the LOS magnetic fields in the photosphere as seen in the Bifrost 100~km simulation (upper left panel). The corresponding emissions in the chromospheric \ion{Ca}{2}~8542~\AA\/ and \ion{Mg}{2}~k$_{3}$ lines are shown in the upper right and bottom left panels, respectively. These images indicate that the stronger Bifrost bipole generates significant activity in the chromospheric layers. Emission in the core of the \ion{Ca}{2} line is notably visible above the footpoints, along magnetic field lines, and along polarity inversion lines. This suggests local heating events occurring above the emerging fields. The emission at the TR heights (\ion{Si}{4}~1393.8~\AA\/, bottom right panel) seems to occur only along the outer edge of the emerging system, due to interaction with the surrounding magnetic fields. 

Further insight may be gained by considering the spectral line profiles of chromospheric and TR lines. Figures~\ref{fig:mg_profiles_bifrost1} and \ref{fig:si_profiles_bifrost1} present the line profiles of the chromospheric \MgIIk\ and TR \SiIV\ lines. The \MgIIk\ line core formed in the upper chromosphere where $\beta \ll 1$, shows strongly enhanced emission in several locations throughout the Bifrost~1 region. Also, the average line core profile in this region is significantly brighter than the average over the computational domain. In isolated locations where the emission is strongest, we find that the \MgIItriplet\ triplet line also goes into emission, indicating a heated lower chromosphere as well \citep{2015ApJ...806...14P}. The FWHM width of the k$_3$ core is also increased in Bifrost~1, as compared to the average over the entire computational domain. However, comparing the line profile to a typical IN region observed by IRIS, we find that the profile is narrower than what is observed, even in typical relatively QS areas, as is evident in Figure~\ref{fig:mg_profiles_bifrost1}, where the average IN spectrum taken on 2014 February 25, starting at 18:59 UT, is shown.
While the \SiIV\ line profiles show locations of strongly enhanced emission in the Bifrost~1 region, the average profile strength and shape are not much different than what are found on average over the entire computational domain. We note that both average profiles are red-shifted by some 5~km/s. However, individual profiles, particularly in the Bifrost~1 region, show greatly enhanced emission, line shifts (both red and blue), and widths. The line shifts can be $ > 50$~km/s, while the widths (FWHM) reach several tens of kilometers per second, comparable to or greater than the speed of sound at the nominal temperature of \SiIV\  ions, which is of order 30~km/s.

It is expected that the newly emerging fields may lead to heating in the upper solar atmosphere through various processes, such as magnetic field reconnection within the emerging bipoles or with the surrounding and the overlying fields. Other possible mechanisms include magnetoacoustic waves, shocks, braiding of magnetic field lines, and swirls. The precise mechanisms driving the heating within the Bifrost bipoles shown in this paper will be examined in our forthcoming paper. In any case, our Bifrost simulation provides examples of emerging fields with varying sizes and strengths whose impact on the chromosphere and TR will be examined in the future.

\begin{figure*}[!t]
	\centering
	\resizebox{1\hsize}{!}{\includegraphics[width=1.0\textwidth]{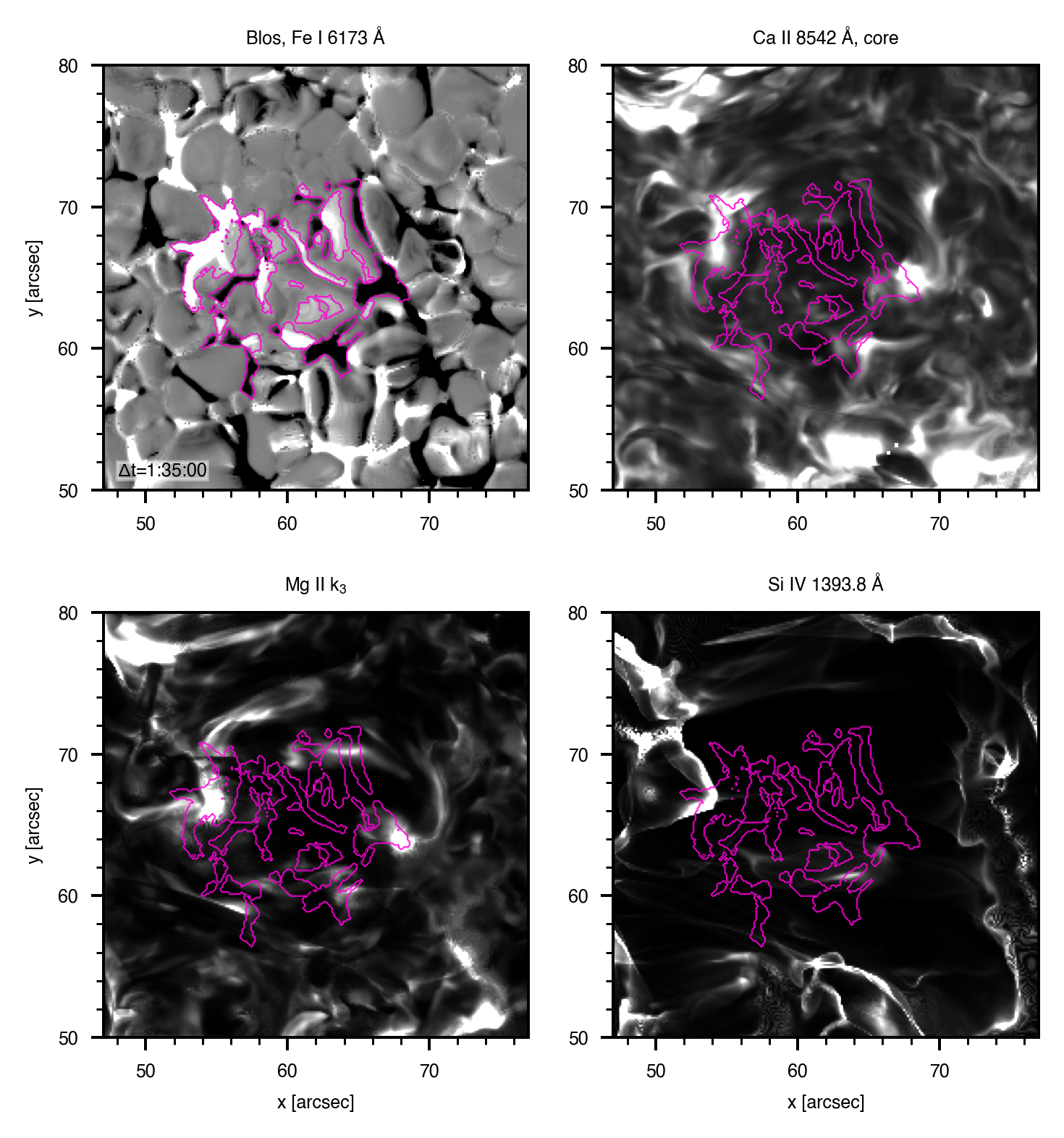}}
	\caption{The same as Figure~\ref{fig:fe_ca_mg_bifrost1} but for the Bifrost 2 bipole.}
	\label{fig:fe_ca_mg_bifrost2}
\end{figure*}

Figure~\ref{fig:fe_ca_mg_bifrost2} illustrates the chromospheric response to the smaller and weaker Bifrost bipole emerging from below. This bipole appears to create a less active environment, in which most of the heating originates from the footpoints. Between the footpoints, the emission level closely resembles the background level. Slightly increased emission in the TR is visible only in the southern part of the emerging bipole, where the opposite-polarity footpoints are canceling.

\begin{figure*}[!t]
	\centering
	\resizebox{1\hsize}{!}{\includegraphics[width=1.0\textwidth]{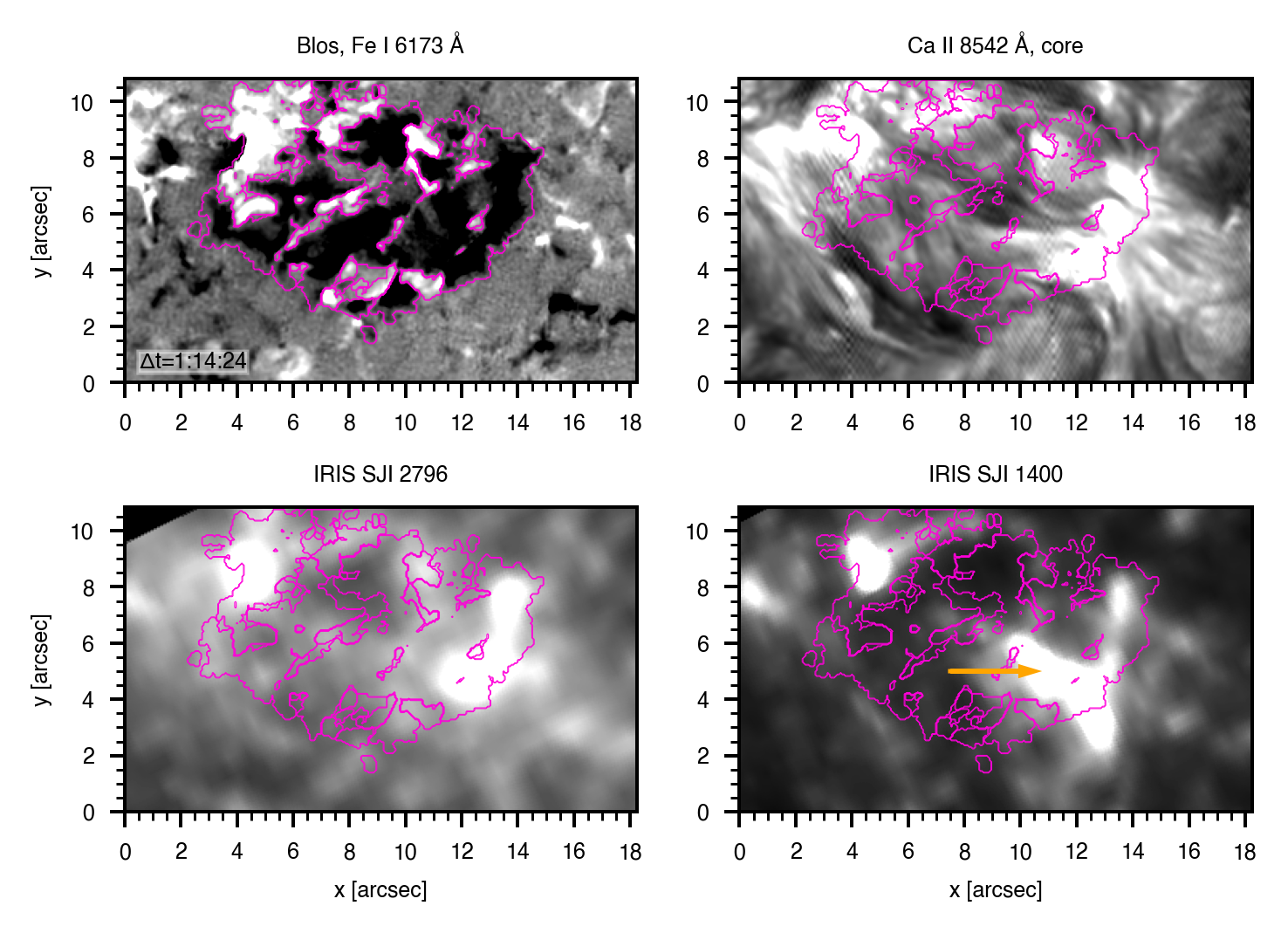}}
	\caption{Upper row: SST LOS magnetic field map in the \ion{Fe}{1} 6173 \AA\/ line and the core intensity filtergram in the \ion{Ca}{2} 8542 \AA\/ line. Bottom row: IRIS SJI 1400 and 2796 intensity maps. The images show the largest emerging bipole detected in the SST15 data set, where many loop-like structures are observed to connect the footpoints and heat the upper solar atmosphere. The orange arrow points at a jet-like eruption.}
	\label{fig:fe_ca_mg_sst15}
\end{figure*}

The SST15 bipole, the largest detected in our SST data sets, exhibits a similar spatiotemporal evolution to that described in \citep{2021ApJ...911...41G}. Throughout its lifetime, the SST15 bipole perturbs the overlying atmosphere, generating numerous chromospheric features that can be interpreted as the acceleration of chromospheric material driven by strong thermal and magnetic pressure gradients during the expansion of the emerging loops. This may result in the formation of cool jets and subsequent shocks in the overlying atmospheric layers \citep{2016ApJ...822...18N, 2017ApJ...850..153N}, as indicated by the jet-like eruption visible in SJI 1440 filtergrams around $(x,y)=(11\arcsec, 5\arcsec)$, and shown with the orange arrow in Figure~\ref{fig:fe_ca_mg_sst15}.

\section{Discussion and Conclusions}
\label{conclusions}

In this study, we have presented two models of QS-like magnetic fields generated using the Bifrost code and compared these models with observations from the SST and IRIS. We have tracked flux features within the available FOVs and determined the physical parameters of these features. Additionally, we have identified the largest bipolar structures in the Bifrost and SST data, as well as one smaller bipole from Bifrost, and examined their properties in the photosphere. A brief description of their impact on the chromospheric and TR layers is also provided.

When considering the entire FOVs of the Bifrost and SST datasets, we found that the detected flux patches in the Bifrost snapshots carry approximately an order of magnitude more flux than those in the SST patches, around $10^{17}$~Mx. This indicates that Bifrost patches represent NW-like magnetic structures, while the SST measurements reveal a combination of weak IN patches and strong NW patches. The consistently stronger fields observed in Bifrost snapshots are also reflected in higher flux densities. As a result, the Bifrost magnetic fields, which are stronger and larger than those detected in the SST data, are more resilient to convective motions and tend to last longer (approximately 4 minutes) compared to SST magnetic elements (whose lifetimes are less than 2 minutes).

The rates of appearance and disappearance of magnetic flux in the SST observations are somewhat higher (approximately 300~Mx~cm$^{-2}$~day$^{-1}$) than in the Bifrost snapshots (approximately 170~Mx~cm$^{-2}$~day$^{-1}$). However, once newly appeared magnetic fields reach the photosphere, they exhibit similar dynamics across the datasets. Specifically, the horizontal velocities measured in all datasets are quite similar, ranging between 2 and 3~km~s$^{-1}$. Additionally, the magnetic elements in all datasets undergo a comparable number of merging and fragmentation events. One might expect the Bifrost magnetic elements, due to their longer lifetimes, to have more surface interactions. Nonetheless, it is important to note that the downhill method used to identify the magnetic patches is sensitive to noise in the SST measurements, which can artificially shorten lifetimes and increase the rates of fragmentation and merging.

Despite the fact that the Bifrost fields are generally larger and stronger than those detected in the SST data, the emerging fields display a closer resemblance. During the emerging phase, all three detected bipoles expand at rates consistent with large QS bipoles, ranging from 0.4 to 3.9~km~s$^{-1}$. The individual magnetic elements forming the bipolar footpoints of the smaller Bifrost bipole and the SST ER have nearly identical flux density distributions. However, within the larger Bifrost bipole, it remains more likely to encounter stronger fields rather than weaker ones. Although the Bifrost patches have larger effective sizes, their total magnetic fluxes are only slightly larger, and less than an order of magnitude different, compared to the values derived from the entire FOV analysis.

In total, the strongest Bifrost bipole contributes $13 \times 10^{19}$~Mx to the solar surface. The smaller Bifrost bipole and the SST ER have more similar flux values, measuring $5 \times 10^{19}$~Mx and $2 \times 10^{19}$~Mx, respectively. At their later evolving stages, all three bipoles first exhibit a plateau of fluxes and sizes, followed by a noticeable cancellation with the surrounding magnetic fields.

The distributions of magnetic field strengths, derived from the SIR inversions of the \ion{Fe}{1}~6173 line profiles, demonstrate that on average the strengths are comparable across the bipoles, ranging from 150~G (Bifrost 2) to 190~G (SST15) and 260~G (Bifrost 1). However, the tails of the Bifrost distributions extend more toward stronger, NW-like fields with kG strengths. Additionally, the mean LOS velocities within the footpoints indicate somewhat stronger flows in the Bifrost bipoles. Finding strong fields in IN regions is not uncommon, but there are indications that higher-resolution observations may reveal more of the kG fields. The recent measurements from the Daniel K. Inouye Solar Telescope \citep[DKIST;][]{2014SPIE.9147E..07E}, showed the presence of about 2~kG fields in the
intergranular lanes, which is the highest field strength ever recorded for the IN \citep{2023ApJ...955L..36C}. It is unknown how common such a strong field is in IN areas, but it is unlikely that the discrepancy between the observations and simulations presented in this work could be resolved only by spatial resolution. Instead, part of the discrepancy likely also comes from stronger input fields in our Bifrost simulations.

After the observed bipoles emerge in the solar photosphere, they continue to rise. Both the larger Bifrost bipole and the SST ER reach the upper chromosphere, as indicated by emissions in the core of the \ion{Ca}{2} 8542~\AA\/ line and the \ion{Mg}{2}~k. Snapshots of \ion{Si}{4}~1393.8~\AA\/ show activity in the TR above the emerging fields and at the boundary between the emerging and preexisting fields. The smaller Bifrost bipole exhibits activity in the chromosphere, particularly along the interface between the emerging and preexisting fields. At heights corresponding to the TR, this smaller bipole appears to produce only weak emissions coinciding with the canceling footpoints.

The two Bifrost models described in this paper are primarily designed to simulate the chromospheric NW and the overlying coronal regions \citep{2020AGUFMSH0010021H}. However, the models extend from the deep convective zone to the outer corona. The models illustrate flux emergence through the photosphere, modified by convective photospheric flows. The main advantage of the simulations presented here is that they demonstrate self-consistent coronal heating through the braiding process as well as a response to the flux emergence through the chromosphere into the corona. The variety of bipoles observed in the simulations allows us to study not only those that reach the chromosphere and the atmospheric layers above, but also those that do not significantly disturb the chromosphere.

To summarize, the largest discrepancies between observations and the models presented are reflected through the distribution of field strengths, the sizes of magnetic flux patches, and in the chromospheric response to the presence of fields. We also note that in our {\tt nw072100} Bifrost model, magnetic features live longer than those identified in the SST measurements, likely due to a lower data cadence in Bifrost compared to the SST datasets. While the observed magnetic fields show a combination of weaker and stronger fields, the Bifrost models have a preponderance of stronger fields in larger patches. The differences in strength and size may be accounted for by increasing the resolution of the model experiments. The photospheric horizontal and LOS velocities and the number of fragmentations and mergings of flux elements seem well reproduced in the models. However, a comparison of the line widths of \MgIIk\ (and \SiIV) shows that the modeled upper chromosphere lacks either vigorous turbulent velocities or sufficient opacity (e.g., mass) to reproduce the observed widths. Again, this difference may be mitigated by running models at higher resolution \citep{2023ApJ...944..131H,2024A&A...692A...6O}, but including additional non-MHD physics may also be required, such as the inclusion of nonequilibrium hydrogen and helium and/or the Generalized Ohm's law.

Our results show that the flux emergence process does increase chromospheric dynamics, energetics, and mass content, leading to larger average widths, but it is far from clear whether this is sufficient to alleviate the differences on large scales. These issues will be the subject of our future work. 

In the follow-up paper, we will describe in detail how Bifrost bipoles emerge through the solar atmosphere. We will also explore the factors that determine the heights the emerging bipoles can reach. 
 
\begin{acknowledgments}
IRIS is a NASA Small Explorer Mission developed and operated by LMSAL, with mission operations executed at NASA Ames Research Center and major contributions to downlink communications funded by ESA and the Norwegian Space Centre. M.G, V.H.H., B.D.P., and A.S.D. are supported by NASA contract NNG09FA40C (IRIS). M.G., V.H.H., and B.D.P. were also supported by NASA grant 80NSSC20K1272. The Swedish 1 m Solar Telescope (SST) is operated on the island of La Palma by the Institute for Solar Physics of Stockholm University in the Spanish Observatory del Roque de los Muchachos of the Instituto de Astrof\'isica de Canarias. 
The SST is co-funded by the Swedish Research Council as a national research infrastructure (registration number 4.3-2021-00169). 
This research is supported by the Research Council of Norway, project number 325491, 
and through its Centres of Excellence scheme, project number 262622. 
This research has made use of NASA's Astrophysics Data System.
\end{acknowledgments}

\bibliographystyle{aasjournal}
\bibliography{sst_bifrost_refs}

\end{document}